 \pdfoutput=1

\documentclass[aps,prd,twocolumn,showpacs, floatfix, letterpaper,preprintnumbers,superscriptaddress,nofootinbib,groupedaddress]{revtex4}

\usepackage{graphicx,epsfig}
\usepackage{amsmath}
\usepackage {amssymb}
\usepackage {longtable}
\usepackage{multirow}
 \usepackage{dcolumn}
\usepackage{bm}
\usepackage{amsfonts}
\usepackage{subfigure}
\usepackage{color}
\usepackage{relsize}
\usepackage{hyperref}
\usepackage{xcolor}
\usepackage{float}
\usepackage{natbib}

\newcommand{\be}{\begin{equation}}
\newcommand{\ee}{\end{equation}}
\newcommand{\bea}{\begin{eqnarray}}
\newcommand{\eea}{\end{eqnarray}}

\begin{document}

\title{Inflation and Primordial Fluctuations in $F(T)$ Gravity's Rainbow}

\author{Yoelsy Leyva}
\email{yoelsy.leyva@academicos.uta.cl }
\affiliation{Departamento de F\'isica, Facultad de Ciencias, Universidad de Tarapac\'a, Casilla 7-D, Arica, Chile}

\author{Carlos Leiva}
\email{cleivas62@gmail.com}
\affiliation{Departamento de F\'isica, Facultad de Ciencias, Universidad de Tarapac\'a, Casilla 7-D, Arica, Chile}

\author{Giovanni Otalora}
\email{giovanni.otalora@pucv.cl}
\affiliation{Instituto de F\'{\i}sica, Pontificia Universidad Cat\'olica de 
Valpara\'{\i}so, 
Casilla 4950, Valpara\'{\i}so, Chile}

\author{Joel Saavedra}
\email{joel.saavedra@pucv.cl}
\affiliation{Instituto de F\'{\i}sica, Pontificia Universidad Cat\'olica de 
Valpara\'{\i}so, 
Casilla 4950, Valpara\'{\i}so, Chile}
\date{\today}

\begin{abstract} 
We study slow-roll inflation and the generation of primordial fluctuations in $F(T)$ gravity's rainbow. We obtain the second order action for scalar and tensor perturbations and then calculate the primordial power spectrum for them. Thus, after calculating the inflationary observables up to first order in slow-roll approximation, namely the scalar spectral index $n_{s}$  and the tensor-to scalar ratio $r$, we confront the predictions of the model with the current PLANCK and BICEP/Keck data. 

\end{abstract}


\maketitle

\section{Introduction}\label{Introduction}
Cosmic inflation is the most successful theory in explaining the evolution of the very early universe. The central idea in this inflationary paradigm is that the primordial universe underwent a quasi-exponentially accelerated phase \cite{Starobinsky,Guth:1980zm,Albrecht:1982wi,Linde:1981mu}. It gives a solution for the well-known theoretical problems of the standard hot Big-Bang cosmological model such as the horizon, the flatness, and the topological defects problems, etc. Furthermore, inflation itself provides a mechanism for the generation of a Gaussian and nearly scale-invariant spectrum of primordial perturbations in agreement with observations \cite{Planck:2015sxf,Planck:2018jri,BICEP2:2015nss,BICEP2:2018kqh}. In cosmic inflation these primordial perturbations can arise naturally from the quantum fluctuations of the inflaton field, a dynamical scalar field $\phi$ minimally coupled to Einstein Gravity and which is slowly rolling down an effective nearly-flat scalar potential $V(\phi)$ \cite{Senatore:2016aui,Riotto:2018pcx,Baumann:2018bq,Baumann:2014nda,Weinberg:2008zzc}. In fact, a generic  prediction of inflation is the production of a stochastic background of primordial gravitational waves (PGWs) and currently many efforts are being made to detect them \cite{Maggiore:2018sht}. The quantity used to parameterize the amplitude of these PGWs is the ratio between the amplitude of primordial tensor perturbations and the amplitude of primordial scalar perturbations, the so-called tensor-to-scalar ratio $r$. Recently, new data from BICEP/Keck $2021$ have been released  \cite{BICEPKeck:2021gln}, improving the constraints on primordial gravitational waves and putting a considerably stronger upper bound on $r$: $r_{0.05}=0.014^{+0.010}_{-0.011}$ ($r_{0.05}<0.036$ at $95\%$ confidence level (C.L.)), when compared with the results from Planck $2018$ \cite{Planck:2018jri,BICEP2:2015nss,BICEP2:2018kqh}: $r_{0.002}<0.064$ at $95\%$ C.L. 

In the geometrical description of General Relativity (GR) the gravitational interaction is described in terms of the curvature associated with the Levi-Civita Connection \cite{Weinberg:1972kfs,Wald:1984rg}. Moreover, it is well-known that gravity can also be described in terms of torsion in the context of the so-called Teleparallel Equivalent of GR, or simply Teleparallel Gravity (TG) \cite{Einstein,TranslationEinstein,Early-papers1,Early-papers2,Early-papers3,Early-papers4,Early-papers5,Early-papers6}. TG is a gauge theory for the translation group in which the dynamical variable is the tetrad field rather than the metric tensor and torsion is associated to the Weitzenb\"{o}ck connection that substitutes the Levi-Civita connection \cite{JGPereira2,AndradeGuillenPereira-00,Arcos:2005ec,Pereira:2019woq}. Additionally, the Lagrangian density of TG becomes proportional to the torsion scalar $T$, which is equivalent to the curvature scalar $R$ up to a total derivative term. Therefore the two theories, GR and TG, are equivalent at the level of field equations.
\cite{Aldrovandi-Pereira-book,Arcos:2005ec}. 

However, we can modify gravity starting from TG instead of GR, and then one obtains a new class of modified gravity theories based on torsion \cite{Cai:2015emx,Bahamonde:2021gfp}. The prototype and most simple model of this class is $F(T)$ gravity \cite{Bengochea:2008gz,Linder:2010py,Li:2011wu}. In this theory the Lagrangian density of TG is promoted to an arbitrary function of $T$ in close analogy to $F(R)$ gravity \cite{Clifton:2011jh,Capozziello:2011et,DeFelice:2010aj,Nojiri:2010wj,Nojiri:2006ri}. Furthermore, a self-interacting scalar field gravitating in $F(T)$ gravity with accelerating solutions was first studied in Ref. \cite{Yerzhanov:2010vu} and also later in Ref. \cite{Chakrabarti:2017moe} from the view of point of a reconstruction scheme in an accelerating Universe. Also, a more general scalar-torsion $f(T,\phi)$ can also be proposed \cite{Hohmann:2018rwf,Gonzalez-Espinoza:2020azh}, that includes a wide family of theories such as $F(T)$ gravity plus a scalar field \cite{Yerzhanov:2010vu,Chakrabarti:2017moe,Rezazadeh:2015dza,Goodarzi:2018feh,Bamba:2016gbu},  non-minimally coupled scalar-torsion theories  \cite{Geng:2011aj,Geng:2011ka,Xu:2012jf,Wei:2011yr,Otalora:2013tba,Otalora:2013dsa,Otalora:2014aoa,OtaloraPatino:2014aru,Skugoreva:2014ena,Jarv:2015odu,Gonzalez-Espinoza:2019ajd}, and extensions of this latter through a non-linear scalar-torsion coupling \cite{Gonzalez-Espinoza:2020azh,Gonzalez-Espinoza:2020jss,Gonzalez-Espinoza:2021qnv,Gonzalez-Espinoza:2021mwr}. 

Gravity's Rainbow is an attempt to realize a small-scale, Ultra-Violet (UV) modification of GR, and such that GR is recovered as a low-energy limit \cite{Magueijo:2002xx}. As in other approaches to quantum gravity, this theory arises from the idea of overcoming the non-renormalizability of GR and the difficulties encountered when trying to quantize gravity \cite{Stelle:1976gc}. Unlike what happens for instance in Ho\v{r}ava-Lifshitz gravity \cite{Horava:2009uw} where the UV modification of GR is proposed from the gravitational action itself, in Rainbow Gravity the UV modification is postulated from the spacetime metric, and yet keeping a connection with the former since both are Lorentz violating theories \cite{Garattini:2014rwa}. It is constructed as an extension of double special relativity to curved spacetimes, that modifies the principles of special relativity by introducing a new universal constant, the Planck energy, besides the speed of the light. As a result of this, the usual energy-momentum relations are modified due to the emergence of new contributions that depend on the probe energy. So, in order to deal with the fact that the dual position space is non-trivial because of the non-linearity of the Lorentz transformations in the momentum space, one could follow the path of assuming a modified spacetime geometry that depends on the probe energy. Thus, in Rainbow Gravity the metric tensor is deformed near the Planck scale with its components being functions of the energy of the particle probing the spacetime, and the standard energy-independent metric is recovered at low energies \cite{Magueijo:2002xx}.

Inflation from curvature-based modified gravity models under the presence of rainbow gravity effects has already been studied in the past. For instance, in Ref. \cite{Chatrabhuti:2015mws} (see also Ref. \cite{Channuie:2019kus}) the authors studied the Starobinsky model of inflation in the context of rainbow gravity. They proposed that the rainbow functions can be written as power-law functions of the Hubble rate, and then obtained the corresponding modified power spectrum of scalar and tensor perturbations which were contrasted with observations through the scalar spectral index $n_{s}$ and the tensor-to-scalar ratio $r$. Later in Ref. \cite{Waeming:2020rir} the previous results were extended by including in the analysis other inflationary models besides the Starobinsky one, such as the logarithmic-corrected $R^2$ model and the Einstein-Hu-Sawicki model. Furthermore, the authors performed a comparison with the latest Planck 2018 data, concluding a good agreement between the predictions of these models and the observations. 

We would like to highlight that the study of inflation from torsion-based modified gravity models with rainbow gravity effects has not been done so far. Here we intend to fill this gap. It could constitute a very interesting path to understand the physics of inflation since the torsion-based modified gravity represents a good alternative to its counterpart based on curvature \cite{Cai:2015emx}. For instance, the study of inflation and the generation of primordial fluctuations in $f(T,\phi)$ gravity without rainbow gravity effects was performed in Ref. \cite{Gonzalez-Espinoza:2020azh}. Also, a reconstruction scheme for inflation through the parametrization (or attractor) of $n_{s}$ and $r$ in the framework of (non-rainbow) $f(T,\phi)$ gravity was carried out in Ref. \cite{Gonzalez-Espinoza:2021qnv}. 

In the present paper we are interested in studying inflation and the generation of primordial fluctuations in $F(T)$ gravity, taking into account the semiclassical effects of relativistic particles that are probing the spacetime in the very early universe in the same way as in Rainbow Gravity. In this $F(T)$ gravity's rainbow, and as a natural choice when investigating the evolution of the very early universe, we consider for the matter content a canonical scalar field representing the inflaton. Although this scalar field can be explicitly coupled to gravity \cite{Faraoni:2000wk}, for simplicity we are going to assume only the minimal case.

The manuscript is organized as follows: in Section \ref{GRainbow}, we review the basics of Gravity's Rainbow. In Section \ref{FTgravityRainbow} we introduce the $F(T)$ Gravity's Rainbow theory in the inflationary context. In Section \ref{Setup_Slow_Roll} we establish the general setup for slow-roll inflation. In Section \ref{Second_Order}  we study primordial fluctuations by expanding the action up to second order and then we obtain the scalar and tensor power spectrum. In Section \ref{Concrete_Models} we study a concrete inflationary model and confront its predictions with the current PLANCK and BICEP/Keck data. Finally, Section \ref{conclusion_f} is devoted to the conclusions.


\section{Gravity's Rainbow}\label{GRainbow}

With the idea to construct a semiclassical or effective theory of quantum gravity where besides the speed of the light $c$, one more fundamental and universal constant (the same for all the inertial observers) is present, i.e., the Planck length or Planck energy, Magueijo and Smolin \cite{Magueijo:2002xx} proposed to extend doubly special relativity to curved spacetimes. In this latter theory, due to the non-linearity of Lorentz transformations in the momentum space, the invariant of energy and momentum leads to the modified dispersion relation
\be
\mathcal{E}^2 \tilde{f}^2(\mathcal{E})-\mathbb{P}^2 \tilde{g}^2(\mathcal{E})=\mathit{m}_{0}^2,
\label{MDR}
\ee where $\tilde{f}(\mathcal{E})$ and $\tilde{g}(\mathcal{E})$ are functions of energy $\mathcal{E}$. Also, $\mathbb{P}$ is the momentum vector of the probe particle. To give a definition of the position space based on the requirement of linearity in the contraction between position and momentum, one is led to the fact that this position space is endowed with an energy dependent invariant. Indeed, analyzing the null geodesics for light rays one finds that the speed of the light is $c(\mathcal{E})=\tilde{g}(\mathcal{E})/\tilde{f}(\mathcal{E})$. Thus, the modified equivalence principle implies that the spacetime geometry is described by an effective energy-dependent rainbow metric
\be
g(\mathcal{E})=\eta_{A B} e^{A}(\mathcal{E}) \otimes e^{B}(\mathcal{E}),
\ee where $\eta_{A B}=\text{diag}\,(-1,1,1,1)$ is the Minkowski tangent space metric, and the tetrad fields are now energy dependent in the way
\be
e^{\hat{0}}=\frac{1}{\tilde{f}^2(\mathcal{E})} \bar{e}^{\hat{0}},\:\:\: e^{\hat{i}}=\frac{1}{\tilde{g}^2(\mathcal{E})} \bar{e}^{\hat{i}},
\ee where $\bar{e}^{A}$ are the original frame fields and by using the correspondence principle, the rainbow functions $\tilde{f}$ and $\tilde{g}$ tend to unit for $\mathcal{E}/\mathcal{E}_{pl}\ll 1 $, with $\mathcal{E}_{pl}$ the Planck energy. Thus, in this rainbow background, the Lorentz connection, the curvature tensor and the Einstein field equations become energy dependent \cite{Feng:2017gms, Dehghani:2020jcw,Dehghani:2021civ}. 

To apply in the context of cosmology we choose the homogeneous and isotropic background 
\be
e^A_{~\mu}={\rm diag}\left(1/\tilde{f},a/\tilde{g},a/\tilde{g},a/\tilde{g}\right),
\label{RainbowFRW}
\ee which leads to the modified Friedmann-Robertson-Walker (FRW) metric \cite{Ling:2006az}
\bea
ds^2=-\frac{1}{\tilde{f}^2} dt^2+\frac{a^2}{\tilde{g}^2} \delta_{i j}dx^{i} dx^{j},
\eea where $a(t)$ is the scale factor of the Universe which is a function of the cosmic time $t$.  At the limit $\tilde{f}, \tilde{g}\rightarrow 1$ we recover the standard FRW background. Starting from this rainbow background the corresponding modified Friedmann equations can be calculated. Furthermore, it is very reasonable to think that in an expanding universe the probe energy may be time dependent and then also the rainbow functions via the expansion Hubble rate $H\equiv \frac{1}{a} \frac{da}{dt}$ \cite{Ling:2006az,Chatrabhuti:2015mws,Waeming:2020rir}. Below we study the implications of this rainbow background for inflation in the context of $F(T)$ gravity \cite{Cai:2015emx}.

\section{$F(T)$ Gravity's Rainbow}\label{FTgravityRainbow}

The relevant action is given by 
\be
S=\int{d^{4}x e F(T)}+\int{d^{4} x e \mathcal{L}_{m}},
\label{action}
\ee where $F(T)$ is a function of the torsion scalar $T$ and $\mathcal{L}_{m}$ is the Lagrangian density of matter \cite{Cai:2015emx}.
We assume for the matter sector the Lagrangian density of a scalar field
\be
\mathcal{L}_{m}=P(\phi)X-V(\phi),
\label{L_matter}
\ee where $X\equiv -\partial_{\mu}{\phi}\partial^{\mu}{\phi}/2$ and $V(\phi)$ is the scalar potential. $P(\phi)$ is a function of the scalar field and for $P(\phi)=1$ we recover the Lagrangian density of a canonical scalar field  \cite{Weinberg:2008zzc}.

Varying the action with respect to the tetrad field $e^{A}_{~\mu}$ we obtain the corresponding modified field equations
\begin{eqnarray}
&& \frac{1}{e}\partial_{\mu}\left(e F_{T} e_{A}^{~\tau} S_{\tau}^{~\rho \mu}\right)-F_{T} 
e_{A}^{~\tau} S_{\nu}^{~\mu \rho} T^{\nu}_{~\mu \tau}+\frac{1}{4} 
e_{A}^{~\rho}F\nonumber\\
&& =\frac{1}{4} e_{A}^{~\tau}\,
  {\mathcal{T}^{(m)}}_{\tau}^{~\rho},
\label{FEquations}
\end{eqnarray}
where we have defined the ``superpotential'' $S_{\rho}^{~\mu\nu}\equiv 
\frac{1}{2}\left(K^{\mu\nu}_{~~\rho}+\delta^{\mu}_{\rho}\,T^{\theta\nu}_{
~~\theta}-\delta^{\nu}_{\rho}\,T^{\theta\mu}_{~~\theta}\right)$, with 
$K^{\mu\nu}_{~~\rho}\equiv 
-\frac{1}{2}\left(T^{\mu\nu}_{~~\rho}-T^{\nu\mu}_{~~\rho}-T_{\rho}^{~\mu\nu}\right)$   
the contortion tensor. Also, the matter energy-momentum tensor is defined as
\be
 {\mathcal{T}^{(m)}}_{\mu}^{~\nu}\equiv e^{A}_{~\mu}\left[\frac{1}{e}\frac{\delta S_{m}}{\delta e^{A}_{~\nu}}\right], 
\ee where $S_{m}=\int{d^{4}x e \mathcal{L}_{m}}$ is the action of the matter field and $\mathcal{L}_{m}$ is given by Eq. \eqref{L_matter}. 

By assuming the rainbow background \eqref{RainbowFRW} we find the following modified cosmological equations
\bea
\label{FRW00}
&& F- 2 T F_{,T}=\frac{1}{2} P \tilde{f}^2 \dot{\phi}^2+V,\\
&&\dot{T}\left[ 2 T F_{,TT}+F_{,T}\right]=\frac{3 P \tilde{f}^2 \left(H \tilde{g}- \dot{\tilde{g}}\right)\dot{\phi}^2}{\tilde{g}},
\label{FRWii}
\eea where the torsion scalar $T$ is now given by
\be
T=\frac{6\tilde{f}^2 \left(H \tilde{g}- \dot{\tilde{g}}\right)^2}{\tilde{g}^2}.
\label{T_R}
\ee

Also, for the scalar field we obtain the modified motion equation
\be
\ddot{\phi}+ \left[3 H+\frac{\dot{P}}{2P}+ \left(\frac{\dot{\tilde{f}}}{\tilde{f}}-\frac{3 \dot{\tilde{g}}}{\tilde{g}}\right)\right]\dot{\phi}+\frac{V_{,\phi}}{\tilde{f}^2 P}=0.
\label{Motion_Eq}
\ee Notice that the rainbow functions contribute to the friction term. The comma means derivative with respect to $T$, or $\phi$, and the dot derivative with respect to the cosmic time $t$. 

Below we study slow-roll inflation in this cosmological framework.




\section{General setup for slow-roll inflation}\label{Setup_Slow_Roll}

Slow-roll inflation is currently the most successful theory in explaining the physics of the very early universe \cite{Weinberg:2008zzc,MukhanovBook}. In order to study slow-roll inflation we introduce the following set of dimensionless parameters
\bea
&& \epsilon=-\frac{\dot{H}}{H^2},\:\:\: \delta_{PX}=-\frac{3 P \tilde{f}^2 X}{T F_{,T}},\:\:\:\: \delta_{F_{,T}}=\frac{\dot{F}_{,T}}{H F_{,T}},\nonumber\\
&& \delta_{\tilde{f}}=\frac{\dot{\tilde{f}}}{H f},\:\:\:\: \delta_{\tilde{g}}=\frac{\dot{\tilde{g}}}{H g},\:\:\: \eta_{\tilde{f}}=\frac{\dot{\delta}_{\tilde{f}}}{H \delta_{\tilde{f}}},\:\:\: \eta_{\tilde{g}}=\frac{\dot{\delta}_{\tilde{g}}}{H \delta_{\tilde{g}}},\nonumber\\
&& \delta_{\phi}=\frac{\ddot{\phi}}{H \dot{\phi}},\:\:\: \delta_{P}=\frac{\dot{P}}{H P},\:\:\: \eta_{F_{,T}}=\frac{\dot{\delta}_{F_{,T}}}{H\delta_{F_{,T}}}.
\label{Slow_Para}
\eea Since $\epsilon\ll 1$ during inflation, all the parameters defined in Eq. \eqref{Slow_Para} are much smaller than the order of the unity. 

The equation \eqref{FRW00} can be wrriten as
\be
\frac{F}{T F_{,T}}-2=-\frac{1}{3} \delta_{PX}+\frac{V}{T F_{,T}}, 
\ee and then at leading order we obtain
\be
F-2 T F_{,T}\simeq V.
\label{slow1}
\ee In Eq. \eqref{slow1} we have applied the slow-roll approximation \cite{Weinberg:2008zzc,MukhanovBook} in the context of modified teleparallel gravity theories (see also Refs. \cite{Gonzalez-Espinoza:2020azh,Gonzalez-Espinoza:2021qnv}). For $F(T)=-T/(2 \kappa^2)$ we recover the usual relation $3 H^2/\kappa^2 \simeq V$ that occurs for single-field slow-roll inflation \cite{Weinberg:2008zzc,MukhanovBook,Baumann:2014nda}.  Similarly, the motion equation \eqref{Motion_Eq} gives
\be
\frac{1}{3} \delta_{\phi}+\frac{1}{6}\delta_{P}+\frac{1}{3} \delta_{\tilde{f}}-\delta_{\tilde{g}}+1+\frac{V_{,\phi}}{3 \tilde{f}^2 P H \dot{\phi}}=0, 
\ee and then at leading order we get
\be
\dot{\phi}\simeq -\frac{V_{,\phi}}{3 P \tilde{f}^2 H}.
\label{dotphi}
\ee
From Eq. \eqref{T_R}, we can write
\be
T=6 \tilde{f}^2 H^2 \left(1-\delta_{\tilde{g}}\right)^2,
\ee and thus, Eq. \eqref{dotphi} yields
\be
\frac{\dot{\phi}}{M_{pl} H}\simeq -\frac{2 V_{,\phi}}{M_{pl} P T}.
\ee In this latter equation, the right hand side is a function of $\phi$. For some ansatzes of the $F(T)$ function, and for a smooth potential $V(\phi)$, we can solve \eqref{slow1} for $T=T(\phi)$.

Also, the equation \eqref{FRWii} leads us to
\be
\delta_{F_{,T}}+\frac{\dot{T}}{2 T H}=-\delta_{PX}+\delta_{PX} \delta_{\tilde{g}}.
\label{SlowEq1}
\ee By using Eq. \eqref{T_R}, the second term in the previous equation can be written as
\be
\frac{\dot{T}}{2 H T}=\delta_{\tilde{f}}-\epsilon+\frac{\delta_{\tilde{g}}\eta_{ \tilde{g}}}{\delta_{\tilde{g}}-1}.
\label{dotT}
\ee Equation \eqref{dotT} gives (up to a factor $2$) the fractional change of the torsion scalar $T$ (Eq. \eqref{T_R}) per Hubble time in terms of the slow-roll parameters that were introduced in Eq. \eqref{Slow_Para}. So, after substituting \eqref{dotT} into Eq. \eqref{SlowEq1} we obtain
\be
\epsilon=\delta_{PX}+\delta_{F,_{T}}+\delta_{\tilde{f}}-\delta_{PX}\delta_{\tilde{g}}+\frac{\delta_{\tilde{g}}\eta_{ \tilde{g}}}{\delta_{\tilde{g}}-1}.
\label{epsilon}
\ee Therefore, at first order, Eq. \eqref{dotT} reduces to
\be
\frac{\dot{T}}{2 H T}\simeq \delta_{\tilde{f}}-\epsilon,
\label{dotT_2}
\ee and then
\be
\epsilon\simeq \delta_{PX}+\delta_{F,_{T}}+\delta_{\tilde{f}}.
\label{epsilon_slow}
\ee 

The number of $e$-folds $N$ measuring the amount of inflation between the time at which the scales of cosmological interest cross the Hubble horizon, $t_{*}$, and the end of inflation, $t_{\text{end}}$, is calculated as
\bea
&& N=\int_{t_{*}}^{t_{\text{end}}}{H dt}=\int_{\phi_{*}}^{\phi_{\text{end}}}{\left(\frac{\dot{\phi}}{M_{pl} H}\right)^{-1}\left(\frac{d \phi}{M_{pl}}\right)},\nonumber\\
&& \simeq \int_{\phi_{\text{end}}}^{\phi_{*}}{\left(\frac{2 V_{,\phi}}{M_{pl} P(\phi) T(\phi)}\right)^{-1}\left(\frac{d \phi}{M_{pl}}\right)}.
\label{N_phi}
\eea
The value of the inflaton field at the end of inflation $\phi_{\text{end}}$ is obtained from Eq. \eqref{epsilon_slow} by solving the equation $\epsilon(\phi_{\text{end}})\simeq 1$. Thus, from Eq. \eqref{N_phi}, and for some special cases, we can solve analytically for $\phi_{*}=\phi(N_{*})$. In other cases, we need to solve numerically \cite{Gonzalez-Espinoza:2019ajd}. 

Below we study the evolution of the cosmological perturbations around the background in Eq. \eqref{RainbowFRW}. 

\section{Primordial Fluctuations}\label{Second_Order}
\subsection{Scalar Perturbations}
\subsubsection{Second Order Action}

The starting point is the Arnowitt-Deser-Misner (ADM) decomposition of the tetrad field  \cite{Wu:2011kh}
\bea
&& e^{0}_{~\mu}=\left(\mathcal{N},\textbf{0}\right),\:\:\:\: e^{a}_{~\mu}=\left(\mathcal{N}^{a},h^{a}_{~i}\right)\label{ADM1},\\
&& e_{0}^{~\mu}=\left(1/\mathcal{N},-\mathcal{N}^{i}/\mathcal{N}\right),\:\:\:\: e_{a}^{~\mu}=\left(0, h_{a}^{~i}\right)\label{ADM2},
\eea where $\mathcal{N}$ is the lapse function, $\mathcal{N}^{i}=h_{a}^{~i}  \mathcal{N}^{a}$ the shift vector, and $h^{a}_{~i}$ the induced tetrad field that satisfies the usual orthogonality conditions $h^{a}_{~j} h_{a}^{~i}=\delta^{i}_{j}$, $h^{a}_{~i} h_{b}^{~i}=\delta^{a}_{b}$, and it is related to the induced metric of the 3-surface through $h_{i j}=\eta_{a b} h^{a}_{~i} h^{b}_{~j}$.

In the uniform field gauge we have $\delta{\phi=0}$, and then we choose
\be
\mathcal{N}=\frac{1+\alpha}{\tilde{f}},\:\:\:\: \mathcal{N}^{a}=\frac{a^{-1} e^{-\mathcal{R}}\delta^{a}_{~i} \partial^{i}{\psi}}{{\tilde{f}}},\:\:\:\: h^{a}_{~i}=\frac{a e^{\mathcal{R}}\delta^{a}_{~j}\delta^{j}_{~i}}{\tilde{g}}.
\label{Uniform_Field_Gauge}
\ee This tetrad field leads to the perturbed metric with rainbow effects 
\bea
 ds^2 &=&-\frac{1}{\tilde{f}^2}\left[\left(1+\alpha\right)^2-a^{-2}e^{-2\mathcal{R}}\left(\partial \psi\right)^2\right]dt^2 \nonumber \\ 
&& +\frac{2\partial_{i}{\psi}}{\tilde{f}\tilde{g}} dt dx^{i}+ \frac{a^2 e^{2\mathcal{R}}}{\tilde{g}^2}\delta_{i j} dx^{i} dx^{j}.
\eea In the case when $\tilde{f}(\mathcal{E})=\tilde{g}(\mathcal{E})=1$ we recover the usual perturbed metric in the uniform field gauge \cite{DeFelice:2011uc}.

The action \eqref{action} is not local Lorentz invariant, and this is reflected in the fact that the field equations \eqref{FEquations} are not symmetric 
 \cite{Sotiriou:2010mv,Li:2010cg}. In order to take into account the effects of the local Lorentz symmetry breaking in modified teleparallel gravity (MTG) we introduce the corresponding additional degrees of freedom through the Lorentz rotation
\be
\Lambda^{A}_{~B}=\left(e^{\chi}\right)^{A}_{~B}=\delta^{A}_{~B}+\chi^{A}_{~B}+\frac{1}{2} \chi^{A}_{~C} \chi^{C}_{~B}+\mathcal{O}(\chi^3).
\label{Lorentz_Transf}
\ee  Then by keeping fixed the vanishing spin connection of the background, we obtain the full perturbed tetrad field
\bea
 e'^{A}_{~\mu}&=&\left(e^{\chi}\right)^{A}_{~B} e^{B}_{~\mu},\nonumber\\
 &=&e^{A}_{~\mu}+\chi^{A}_{~B}e^{B}_{~\mu}+\frac{1}{2} \chi^{A}_{~C} \chi^{C}_{~B} e^{B}_{~\mu}+\mathcal{O}(\chi^3),
\label{Transf_tetrad}
\eea
where the matrix $\chi_{A B}=-\chi_{B A}$ can be parameterized as 
\be
\chi^{0}_{~B}=\left(0,\chi_{b}\right),\:\:\:\: \chi^{a}_{~B}=\left(\chi^{a}, B^{a}_{~b}\right),
\ee such that $\chi^{a}=\eta^{a b}\chi_{b}$ and  $B_{ab}=-B_{ba}$. Thus, one can define the spatial vector $\chi^{i}=h_{a}^{~i} \chi^{a}=\partial_{i}{\beta}+\chi^{(T)}_{i}$, and the spatial antisymmetric tensor $B_{i j}=h^{a}_{~i} h^{b}_{~j} B_{a b}=-B_{j i}=-\epsilon_{j i k} B^{k}$. Hence, there are a scalar mode $\beta$, a transverse vector mode $\chi^{(T)}_{i}$ and a (pseudo) vector mode $B_{i}$ \cite{Wu:2016dkt,Golovnev:2018wbh}.

By following Ref. \cite{Maldacena:2002vr} we expand the action \eqref{action} up to second order in scalar perturbations which gives 
\small
\begin{equation}
\begin{array}{lll}
S^{(2)}&=& \mathlarger{\int} dt d^{3}x a^3\left[ \dfrac{2}{a^2} ( w_1 \dot{\mathcal{R}} - w_2 H \alpha ) \partial^2 \psi + 6 w_3 H \alpha \dot{\mathcal{R}}\right. 
\\
&& -\dfrac{2 w_4 }{a^2} \alpha \partial^2 \mathcal{R} + w_5 \alpha^2 - 3 w_6 \dot{\mathcal{R}}^2 + \dfrac{w_4}{a^2}(\partial \mathcal{R})^2  \\
&& - 4 \left(   w_7 \dot{\mathcal{R}}  - w_8 H \alpha \right) \partial^2 \beta +w_9 \mathcal{R} \partial^2 \beta +  w_{10} (\partial^2 \beta)^{2}   \\
&& \left. + \dfrac{w_{10}}{a^{4}} (\partial^2 \psi)^{2} - \dfrac{2 w_{10}}{a^2} (\partial^2 \beta \partial^2 \psi)  \right],  
\label{second_order}
\end{array}
\end{equation}
\normalsize
where we have defined the functions
\begin{eqnarray}
w_1&=-\frac{2}{\tilde{g}^2}\left(F_{,T}+2 T F_{,T}\right)& , \nonumber
\\
w_{2}&=-\frac{2 \left(H \tilde{g}-\dot{\tilde{g}}\right)}{H \tilde{g}^3} \left(F_{,T}+2 T F_{,TT}\right), \nonumber
\\
w_{3}&=-\frac{2 \tilde{f} \left(H \tilde{g}-\dot{\tilde{g}}\right)}{H \tilde{g}^4} \left(F_{,T}+2 T F_{,TT}\right),\nonumber
\\
w_4&=-\frac{2 F_{,T}}{\tilde{f} \tilde{g}}&  ,\nonumber
\\
w_5&=\frac{P(\phi) \tilde{f}^2 X+ T \left(F_{,T}+2 T F_{,TT}\right)}{\tilde{f} \tilde{g}^3}& ,  \nonumber
\\
w_6 &=-\frac{2 \tilde{f}}{\tilde{g}^3}\left(F_{,T}+2 T F_{,T}\right)&, \nonumber
\\
w_7 &=-\frac{2}{\tilde{g}^2}\left(F_{,T}+T F_{,T}\right)& , \nonumber
\\
w_8 &=-\frac{2 \left(H \tilde{g}-\dot{\tilde{g}}\right)}{H \tilde{g}^3} \left(F_{,T}+T F_{,TT}\right)& ,\nonumber
\\
w_9 &=\frac{24 P(\phi) \tilde{f}^2 X F_{,TT} \left(H \tilde{g}-\dot{\tilde{g}}\right)}{\tilde{g}^3 \left(F_{,T}+2 T F_{,TT}\right)}& ,\nonumber
\\
w_{10} &=\frac{4 T F_{,TT}}{3 \tilde{f}\tilde{g}}&.
\end{eqnarray}

Varying this action with respect to $\partial^{2}\psi$ leads us to
\begin{eqnarray}
w_{1}\dot{\mathcal{R}}-H w_{2}\alpha+ \frac{w_{10}}{a^2} \partial^{2}{\psi}-w_{10} \partial^{2}{\beta}= 0,
\label{var1} 
\end{eqnarray}
whereas variation with respect to $\partial^{2} \beta$ gives
\begin{eqnarray}
&& 2 w_{7}\dot{\mathcal{R}}-2 H w_{8} \alpha -\frac{1}{2} w_{9} \mathcal{R} +\frac{w_{10}}{a^2}\partial^2\psi-\nonumber\\
&& w_{10} \partial^2\beta= 0, 
\label{var2}
\end{eqnarray}
and for $\alpha$ we have
\begin{eqnarray}
&& 3 H w_{3} \mathcal{R}+w_{5} \alpha-\frac{w_{4}}{a^2} \partial^2 \mathcal{R} -\frac{H w_{2}}{a^2} \partial^2\psi+\nonumber\\
&& 2 H w_{8} \partial^2\beta= 0 \label{var3}.
\end{eqnarray}

Thus, from this system of three equations we can solve for  $\alpha$, $\partial^2{\psi}$ and  $\partial^2{\beta}$. Substituting these results in Eq. \eqref{second_order}, and after some integrations  by  parts, we obtain the second order action for the curvature fluctuation
\begin{equation}
S^{(2)} = \int dt d^{3}x a^3 Q_{s} \left[ \dot{\mathcal{R}}^{2} - \dfrac{c^{2}_{s}}{a^{2}} (\partial \mathcal{R})^2 -m^{2} \mathcal{R}^2 \right],  \label{SecOrderSM}
\end{equation}

where
\small
\begin{eqnarray}
\label{Qs}
&& Q_s=\frac{6 P \tilde{f}^3 X}{T \tilde{g}^3}, \\
\label{cs}
&& c^{2}_{s}=\frac{\tilde{g}^2}{\tilde{f}^2},
\\
&& m^2=H^2 \delta_{F_{,T}}\Bigg[6-2\delta_{F_{,T}}+(\delta_{PX}-3)s- \delta_{PX}+\nonumber\\
&& \left(\delta_{PX}-4\right)\delta_{\tilde{f}}+\eta_{F_{,T}}+\eta+\eta_{\tilde{g}}\left(\frac{1}{1-s-\delta_{\tilde{f}}}-1\right)\Bigg].
\label{m2}
\end{eqnarray}
\normalsize
In this latter equation we have defined
\bea
&& s=\frac{\dot{c}_{s}}{H c_{s}}=\delta_{\tilde{g}}-\delta_{\tilde{f}},\\
&& \eta=\frac{\dot{Q}_{s}}{H Q_{s}}=2 \delta_{F_{,T}}+\delta_{P}-s \left(3+2\delta_{PX}\right)+2\delta_{\phi}+\nonumber\\
&& 2 \delta_{PX}\left(1-\delta_{\tilde{f}}\right).
\eea The conditions for the absence of ghost and Laplacian instabilities are written as $Q_{s}>0$ and $c_{s}^2>0$, respectively \cite{DeFelice:2011uc}. From the former condition we require $P>0$ for $\tilde{f},\tilde{g}>0$. Therefore, at first order we get
\be
\eta=2 \delta_{F_{,T}}+\delta_{P}-3 s+2\delta_{\phi}+2 \delta_{PX},
\ee and then
\be
\eta_{\mathcal{R}}\equiv \frac{m^2}{3 H^2}= 2\delta_{F_{,T}}.
\label{eta_R}
\ee This result is consistent with what was found in Ref. \cite{Gonzalez-Espinoza:2020azh} for $f(T)$ gravity plus scalar field in the absence of a rainbow effect. 

\subsubsection{Mukhanov-Sasaki equation}

The second order action \eqref{SecOrderSM} was written in the ``Einstein Frame", where the standard cosmic time $t$ was used and the propagation speed of the scalar perturbations $c_{s}(t)$ is time dependent. For the purposes of our calculations we can go to the ``Rainbow Frame" where a new unit of time (``sound-horizon" time) is introduced such that the propagation speed of the scalar modes becomes equal to $1$ (speed of light in natural units). That is to say, we replace the standard conformal time $d\tau=a^{-1} dt$ by the sound-horizon time $d\tau_{\text{RF}}=c_{s}(\tau) d\tau=(c_{s}(t)/a)dt$ \cite{Amelino-Camelia:2013wha}. Thus, by introducing the canonically-normalized Mukhanov variable $v=z \mathcal{R}$ and $z^2=2 a^2 Q_{s} c_{s}$, the second order action \eqref{SecOrderSM} can be written as  
\begin{equation}
S^{(2)} = \frac{1}{2}\int d\tau_{\text{RF}} d^{3}x  \left[v'^{2} - (\partial v)^2 -M^{2} v^2 \right],  \label{SecOrderSM_2}
\end{equation} where we have defined the effective mass term
\be
M^{2}=\frac{a^2 m^2}{c_{s}^2}-\frac{z''}{z}. 
\ee This effective mass includes the effects of the usual interaction between $\mathcal{R}$ and the cosmological background, as well as the new contributions coming from the violation of local Lorentz symmetry in MTG. Furthermore, this effective mass term also carries the information that was contained in the original propagation speed $c_s$.

Varying the action \eqref{SecOrderSM_2}, and after transforming to the Fourier space, we obtain the Mukhanov-Sasaki equation
\be
v''_{k}+(k^2+M^2)v_{k}=0.
\label{MS}
\ee
 
At first order in slow-roll one has that $a H/c_{s}\simeq -(1+\epsilon+s)/\tau_{\text{RF}}$ and then $z''/z\simeq (1/\tau_{\text{RF}}^2)(2+3 \epsilon +3 \eta /2+9 s/2)$. Therefore, Eq. \eqref{MS} can be rewritten as 
\be
v''_{k}+\left[k^2-\frac{1}{\tau_{\text{RF}}^2}\left(\tilde{\nu}^2-\frac{1}{4}\right)\right] v_{k}= 0, 
\label{MS2}
\ee where we have defined 
\be
\tilde{\nu}^2=\nu^2-3 \eta_{\mathcal{R}}=\frac{9}{4}+3 \left(\epsilon+\frac{1}{2}\eta+\frac{3}{2} s\right)-3 \eta_{\mathcal{R}}.
\ee 
For $\tilde{\nu}$ constant and real, the general solution to Eq.\eqref{MS2} is given by
\be
v_{k}(\tau_{\text{RF}})=\sqrt{-\tau_{\text{RF}}}\left[C_{1} H_{\tilde{\nu}}^{(1)}(-k \tau_{\text{RF}})+C_{2} H_{\tilde{\nu}}^{(2)}(-k \tau_{\text{RF}})\right], 
\ee where $H_{\tilde{\nu}}^{(1,2)}$ are the Hankel’s functions of first and second kind, respectively \cite{Riotto:2018pcx}. After imposing the Bunch-Davies vacuum $v_{k}(\tau_{\text{RF}})\simeq e^{-i k \tau_{\text{RF}}}/\sqrt{2 k}$ at the ultraviolet regime $-k \tau_{\text{RF}}\gg 1$, and with the help of some identities, we get the solution
\be
v_{k}(\tau_{\text{RF}})=\frac{\pi}{2} e^{i \frac{\pi}{2}(\tilde{\nu}+\frac{1}{2})} \sqrt{-\tau_{\text{RF}}} H_{\tilde{\nu}}^{(1)}(-k \tau_{\text{RF}}).
\ee Thus, on super-horizon scales $-k \tau_{\text{RF}}\ll 1$, and also by using some identities, we can write
\be
v_{k}(\tau_{\text{RF}})=\frac{2^{\tilde{\nu}-\frac{3}{2}}}{\sqrt{2 k}}e^{i\frac{\pi}{2}(\tilde{\nu}-\frac{1}{2})} \frac{\Gamma(\tilde{\nu})}{\Gamma(\frac{3}{2})} (-k \tau_{\text{RF}})^{\frac{1}{2}-\tilde{\nu}},
\ee and then we finally find 
\bea
&& \left|R_{k}\right|=z^{-1}\left|v_{k}\right|\simeq \frac{H}{2 \sqrt{Q_{s}c_{s}^3 k^3}}\left(\frac{\tau_{\text{RF}}}{\tau_{\text{RF}}^{*}}\right)^{\frac{3}{2}-\tilde{\nu}},
\label{Rk1}
\eea where $\tau_{\text{RF}}^{*}\simeq -1/k$ is the value of $\tau_{\text{RF}}$ at the horizon crossing.  Also, taking into account the relations
\bea
&& H\simeq H_{*} \left(\frac{\tau_{\text{RF}}}{\tau_{\text{RF}}^{*}}\right)^{\epsilon}, \:\:\: Q_{s}\simeq Q_{s*} \left(\frac{\tau_{\text{RF}}}{\tau_{\text{RF}}^{*}}\right)^{-\eta},\nonumber\\
&& c_{s}\simeq c_{s*}\left(\frac{\tau_{\text{RF}}}{\tau_{\text{RF}}^{*}}\right)^{-s},
\eea the equation \eqref{Rk1} yields
\bea
&&\left|R_{k}\right|\simeq \frac{H_{*}}{2 \sqrt{Q_{s*}c_{s*}^{3} k^3}} \left(\frac{\tau_{\text{RF}}}{\tau_{\text{RF}}^{*}}\right)^{\eta_{\mathcal{R}}},\nonumber\\
&& \simeq \frac{H_{*}}{2 \sqrt{Q_{s*}c_{s*}^{3} k^3}} \left[1+\eta_{\mathcal{R}}\ln{\left(\frac{\tau_{\text{RF}}}{\tau_{\text{RF}}^{*}}\right)}\right],
\label{Rk2}
\eea where $H_{*}$, $Q_{s*}$ and $c_{s*}$ are evaluated at $\tau_{\text{RF}}=\tau_{\text{RF}}^{*}$. 

Thus, the scalar power spectrum of the curvature fluctuation becomes
\bea
&& \mathcal{P}_{s}(k)\equiv \frac{k^3}{2 \pi^2}\left|\mathcal{R}_{k}\right|^2
\simeq \frac{H_{*}^2}{8 \pi^2 Q_{s*} c_{s*}^3}\left(\frac{\tau_{\text{RF}}}{\tau_{\text{RF}}^{*}}\right)^{2 \eta_{\mathcal{R}}}, \nonumber\\
&& \simeq \frac{H_{*}^2}{8 \pi^2 Q_{s*} c_{s*}^3}\left[1+2 \eta_{\mathcal{R}}\ln\left(\frac{\tau_{\text{RF}}}{\tau_{\text{RF}}^{*}}\right)\right].
\label{Ps}
\eea From Eqs. \eqref{Rk2} and \eqref{Ps} one can conclude that the local Lorentz violation in MTG has as consequence a slight logarithmic time dependence of the curvature perturbation and its power spectrum at super-horizon scales \cite{Gonzalez-Espinoza:2020azh}. Since during inflation $\eta_{\mathcal{R}}\sim \mathcal{O}(\epsilon)$, this time dependence can be neglected if needed \cite{Riotto:2018pcx}. Also, let us remember that $\tau_{\text{RF}}/\tau_{\text{RF}}^{*}\simeq c_{s} k/(a H)$ at leading order. 

Also, the spectral index is calculated as
\bea
&& n_{s}-1\equiv \frac{d \ln{\mathcal{P}_{s}(k)}}{d \ln{k}}\simeq -2 \epsilon-\eta-3 s+2 \eta_{\mathcal{R}},\nonumber\\
&& \simeq -4\delta_{PX}-\delta_{P}-2 \delta_{\tilde{f}}-2 \delta_{\phi}.
\label{ns}
\eea Thus, the mass term $\eta_{\mathcal{R}}$ leads to an additional scale dependence of the primordial scalar power spectrum due to the local Lorentz symmetry breaking in MTG \cite{Gonzalez-Espinoza:2020azh}. Although the rainbow slow-roll parameter $\delta_{\tilde{f}}$ and $\delta_{P}$ appear explicitly in the previous expression of $n_{s}$, they are canceled out by the similar terms coming from $\delta_{\phi}$ after using Eq. \eqref{dotphi}.

Below we calculate the primordial power spectrum for tensor perturbations.
\\
\\
\subsection{Tensor perturbations}

In the uniform field gauge $\delta{\phi}=0$, we choose
\be
\mathcal{N}=\frac{1}{\tilde{f}},\:\:\: \mathcal{N}^{a}=0, \:\:\: h^{a}_{~i}=\frac{a}{\tilde{g}}(\delta^{a}_{~i}+\frac{1}{2}\gamma^{a}_{~i}).
\ee 

Therefore, for the second order action of tensor perturbations, we find \cite{Gonzalez-Espinoza:2019ajd}
\be
S_{T}=\sum_{\lambda} \int{dt d^3x a^3 Q_{T}\left[\dot{h}_{\lambda}^2-\frac{c_{T}^2}{a^2} \left(\partial h_{\lambda}\right)^2\right]}.
\label{Tensor_Modes}
\ee The two corresponding polarization states are denoted by $\lambda=+,\times$. Also, we obtained
\be
Q_{T}=-\frac{\tilde{f} F_{,T}}{2 \tilde{g}^3},
\label{Q_T}
\ee
and the squared tensor propagation speed yielded
\be
c_{T}^2=\frac{\tilde{g}^2}{\tilde{f}^2}. 
\ee From Eq. \eqref{Q_T}, and for $\tilde{f},\tilde{g}>0$, the  non-ghost condition $Q_{T}>0$ requires $F_{,T}<0$. In addition, for $\tilde{f}=\tilde{g}$, we obtain $c_{T}=1$ \cite{Magueijo:2002xx}.  

By following a similar procedure than in the case of scalar perturbations, the power spectrum for tensor perturbations becomes \cite{DeFelice:2011zh}
\be
\mathcal{P}_{T}=\frac{H_{*}^2}{2 \pi^2 Q_{T*}c_{T*}^3}.
\ee 
Thus, the tensor spectral index is 
\bea
&& n_{T}\equiv \frac{d \ln{\mathcal{P}_{T}}}{d \ln{k}}=-2\epsilon-\delta_{F_{,T}}+2 \delta_{\tilde{f}},\nonumber\\
&& =-2\delta_{PX}-3\delta_{F_{,T}}.
\label{nT}
\eea 

The tensor-to-scalar ratio is calculated as
\begin{equation}
r =\dfrac{\mathcal{P}_{T}}{\mathcal{P}_{s}}=  16 \delta_{PX}=16 \left(\epsilon-\delta_{F_{,T}}-\delta_{\tilde{f}}\right).
\label{r}
\end{equation}
The slow-roll parameters $\delta_{F_{,T}}$ and $\delta_{\tilde{f}}$ provide small corrections to $r$. For $\delta_{\tilde{f}}=0$ we recover the results obtained in Ref. \cite{Gonzalez-Espinoza:2020azh}. In what follows we apply the above general results to a concrete model of inflation. 

\section{ Chaotic inflation}\label{Concrete_Models}


We study the chaotic potential
\be
V(\phi)=\lambda \phi^{d},
\label{V_phi}
\ee and 
\be
P(\phi)=\xi \phi^{b}, 
\label{P_phi}
\ee where $\lambda$, $\xi$ are positive constants. 

In order to obtain analytical results we take simplest ansatz 
\be
F(T)=-A T^{n},
\label{f_T_Function}
\ee with $A$ a constant. This model is motivated by quantum corrections to the Einstein Hilbert action as considered in curvature-based modified gravity theories \cite{Clifton:2011jh}. For instance, the earliest model of inflation is the Starobinsky model \cite{Starobinsky} which takes into account one-loop corrections to the gravitational action keeping only the $R^2$ correction.  In this way, generalizations of Starobinsky model have also been studied by assuming the $R^{n}$ correction with $n>1$ \cite{Barrow:1988xh,Barrow:1991hg,Martin:2013tda,Sebastiani:2013eqa,Kehagias:2013mya,Costa:2014lta,Cai:2014bda,Chakravarty:2014yda}. In Eq. \eqref{f_T_Function}, we assumed $n\geq 1$ and $A>0$, where $n=1$ corresponds to the case of GR \cite{Cai:2015emx}.

For the rainbow function $\tilde{f}$ we take the ansatz
\be
\tilde{f}=1+\left(\frac{H}{M}\right)^{\gamma}\simeq \left(\frac{H}{M}\right)^{\gamma}, \label{Rainbow_Function}
\ee such that during inflation $H \gg M$, where $M$ is a constant with dimension of mass, and $\gamma$ is a positive exponent. As we have shown, the rainbow function $\tilde{g}$ does not contribute to the dynamics of inflation up to first order in slow-roll approximation. To ensure $c_{s}=c_{T}=1$ in the Einstein frame, one could assume $\tilde{g}=\tilde{f}$ \cite{Magueijo:2002xx}. 

Thus, from Eq. \eqref{slow1} we find
\be
T(\phi)=\left[\frac{\lambda}{A (2 n-1)}\right]^{\frac{1}{n}} \phi^{\frac{d}{n}}. \label{T_phi}
\ee To calculate the amount of inflation we need to have the expressions for the slow-roll parameters in terms of the scalar field. So, from Eq. \eqref{Slow_Para}, it is straightforward to obtain 
\bea
&& \delta_{PX}(\phi)=\frac{d^2 (2 n-1)^{\frac{1}{n}+1} A^{\frac{1}{n}} \lambda ^{1-\frac{1}{n}} }{n \xi }\phi^{-p},\\
&& \delta_{F_{,T}}(\phi)=\frac{2 d^2 (1-n) \left[A (2 n-1)\right]^{\frac{1}{n}} \lambda ^{\frac{n-1}{n}} }{n \xi } \phi ^{-p},\\
&& \delta_{P}(\phi)=-\frac{2 b d \left[A (2 n-1)\right]^{\frac{1}{n}} \lambda ^{1-\frac{1}{n}}}{\xi }\phi^{-p},\\
&& \delta_{\phi}(\phi)= d \left[A (2 n-1)\right]^{\frac{1}{n}} \lambda ^{\frac{n-1}{n}}\times\nonumber\\
&& \left[\frac{2 (b-d+1)}{\xi }+\frac{(2 \gamma+1) d}{(\gamma+1) n \xi }\right] \phi^{-p},
\eea with $p\equiv \left[(b+2)n+d(1-n)\right]/n$, and since 
\be
\delta_{\tilde{f}}(\phi)=-\gamma \epsilon(\phi),
\ee  by using \eqref{epsilon_slow} we get
\bea
&& \epsilon(\phi)=\frac{1}{1+\gamma}\left(\delta_{PX}+\delta_{F_{,T}}\right),\nonumber\\
&& =\frac{d^2 (2 n-1)^{\frac{1}{n}} A^{\frac{1}{n}} \lambda ^{1-\frac{1}{n}} }{(1+\gamma) n \xi }\phi^{-p},
\eea and then
\be
\delta_{\tilde{f}}(\phi)=-\frac{\gamma d^2 (2 n-1)^{\frac{1}{n}} A^{\frac{1}{n}} \lambda ^{1-\frac{1}{n}} }{(1+\gamma)n \xi }\phi^{-p}.
\ee Therefore, the condition $\epsilon(\phi_{\text{end}})=1$ at the end of inflation yields
\be
\phi_{\text{end}}=\left[\frac{d^2 \lambda^{1-\frac{1}{n}} A^{\frac{1}{n}} (2 n-1)^{\frac{1}{n}}}{(\gamma+1) n \xi }\right]^{\frac{1}{p}}.
\ee Notice that $\phi_{\text{end}}$ depends on the rainbow parameter $\gamma$.

After integrating \eqref{N_phi} we obtain the amount of inflation $N$ and then solve $\phi_{*}=\phi(N)$ which gives
\bea
&& \phi(N)=\phi_{\text{end}} \left[1+\frac{2 (\gamma+1)}{1+p (d-2-b)}N\right]^{p}. 
\label{phi_N}
\eea By using the above solution and from Eq. \eqref{Ps}, the scalar power spectrum at the horizon crossing becomes
\bea
&& \mathcal{P}_{s}= \frac{\left(\frac{1}{d^2}\right)^{\frac{d (n-2)}{(b+2) n+d(1-n)}}\left[(2 n-1) \sigma \right]^{\frac{d-4-2 b}{(b+2) n+d(1-n)}}}{96 \pi ^2 \left[(\gamma+1) n\right]^{\frac{(b+2) n+d (3-2 n)}{(b+2) n+d(1-n)}}} \times \nonumber\\
&& \left[1+\frac{2 (\gamma+1) N}{p (d-2-b)+1}\right]^{p \left[b+d \left(\frac{3}{n}-2\right)+2\right]},
\label{PsModel1}
\eea where we have performed the parameter transformation  
\be
A= \sigma  \xi ^{\frac{d (n-2)}{2 b-d+4}} \lambda ^{\frac{(b+2) \left[p (2 b-d+4)-2\right]}{(2 b-d+4)\left[p (b-d+2)-1\right]}}.
\ee From the latest Planck data one has that the current observational value for the amplitude of primordial scalar perturbations is $\mathcal{P}_{s}=2.141 \times 10^{-9}$ \cite{Planck:2018jri}, and then, Eq. \eqref{PsModel1} allows us to solve $\sigma$ in terms of the remaining parameters. 

Now, by using the solution \eqref{phi_N} we obtain 
\bea
&& \delta_{PX}(N)=\frac{(\gamma+1)\left[p (b+d+2)-1\right]}{1-p(2+b-d)+2 (\gamma+1) N},\\
&& \delta_{F_{,T}}(N)=\frac{2 (\gamma +1) \left[1-(b+2) p\right]}{1-p (2+b-d)+2 (\gamma+1) N},\\
&& \delta_{\phi} (N)=\frac{p (b-d)+2 \gamma (1-p)+1}{1-p(b-d+2)+2 (\gamma+1) N},\\
&& \delta_{\tilde{f}}(N)=\frac{\gamma \left[p (b-d+2)-1\right]}{1-p(b-d+2)+2 (\gamma+1) N}.
\eea
Hence, from Eq. \eqref{eta_R}, the mass term $\eta_{\mathcal{R}}$ becomes
\be
\eta_{\mathcal{R}}(N)=\frac{2 (\gamma +1) \left[1-(b+2) p\right]}{1-p (2+b-d)+2 (\gamma+1) N}.
\ee Notice that $\eta_{\mathcal{R}}$  now depends on the rainbow parameter $\gamma$. Near the Hubble horizon crossing, for $N\gg 1$, one has $\eta_{\mathcal{R}}\simeq \left[1-(b+2)p\right]/N$. After the Hubble horizon crossing, at intermediate times and at the end of inflation, the rainbow effect shows up explicitly and becomes important by increasing the value of $\eta_{\mathcal{R}}$ in the way $\eta_{\mathcal{R}}\sim 2 (\gamma +1) \left[1-(b+2) p\right]/\left[1-p (2+b-d)\right]$. Let us remember that we require $\eta_{\mathcal{R}}=2\delta_{F,T}\ll 1$ in order to have $\epsilon\ll 1$. Thus, this latter condition puts an upper bound on the values of $\gamma$ that depends on the other parameters of the model such as  $b$, $d$ and $n$. 

Therefore, the scale dependence of the scalar power spectrum, Eq. \eqref{ns}, is written as
\be
n_{s}(N)=1+ \frac{2 (\gamma+1) \left[p (2 b+d+4)-1\right]}{p (b-d+2)-1-2 (\gamma+1) N},
\label{n_s_N}
\ee and the tensor-to-scalar ratio, Eq. \eqref{r}, becomes
\be
r(N)=\frac{16 (\gamma+1) \left[p (b+d+2)-1\right]}{2 (\gamma+1) N+p (d-b-2)+1}.
\label{r_N}
\ee For $N\gg 1$ we find $n_{s}(N)\simeq 1 -\left[p (2 b+d+4)-1\right]/N$ and $r\simeq \left[8 p (b+d+2)-1\right]/N$.  

\begin{figure}[htbp]
	\centering
		\includegraphics[width=0.45\textwidth]{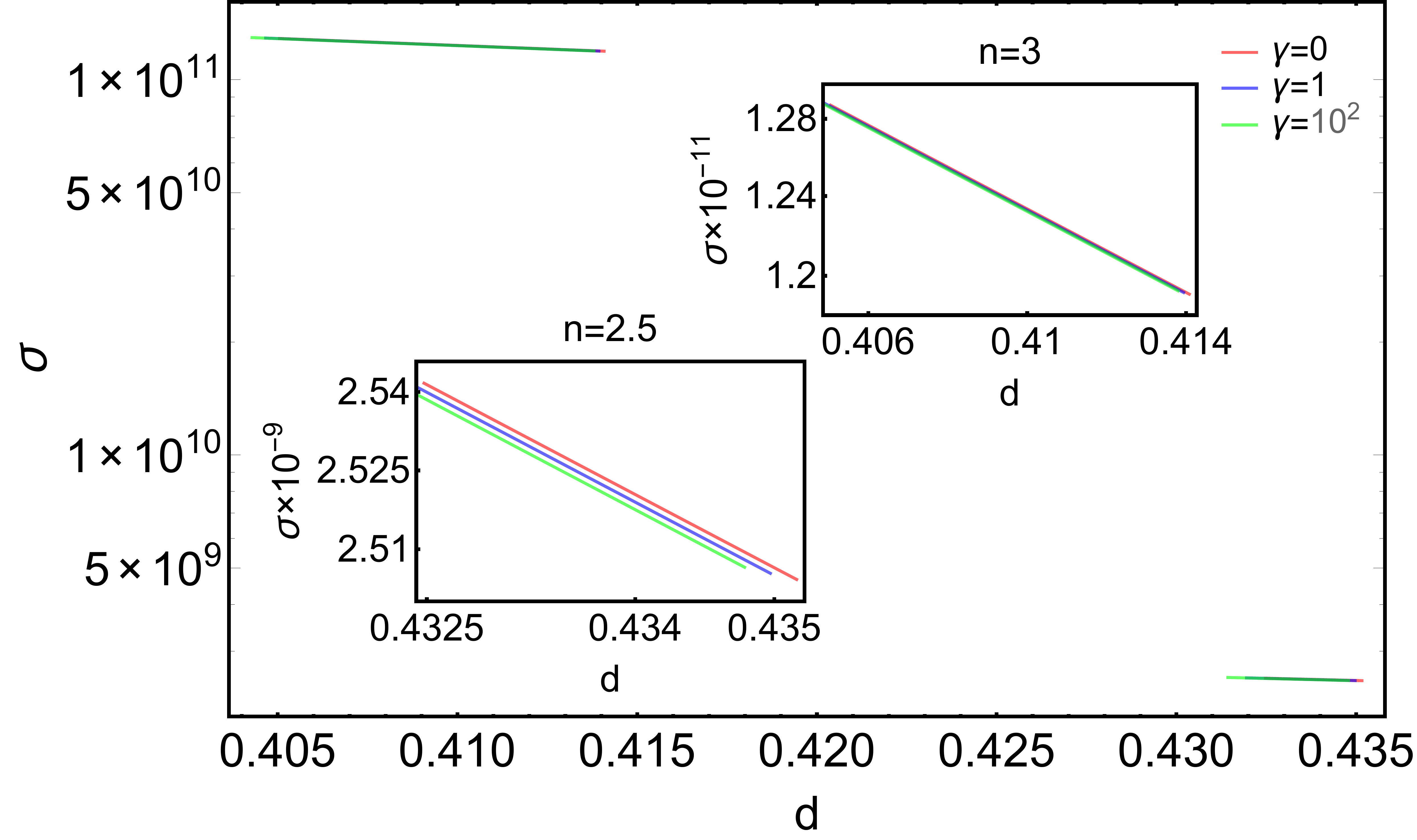}
	\caption{We depict the values of the parameter $\sigma$. We have assumed several different values of the rainbow parameter $\gamma$ and the power exponent $n$. Also, we used $N_{*}=50$ and the current observational value for the amplitude of primordial scalar perturbations $\mathcal{P}_{s}=2.141\times 10^{-9}$ \cite{Planck:2018jri}.}  
	\label{FIG1}
\end{figure}

\begin{figure}[htbp]
	\centering
		\includegraphics[width=0.45\textwidth]{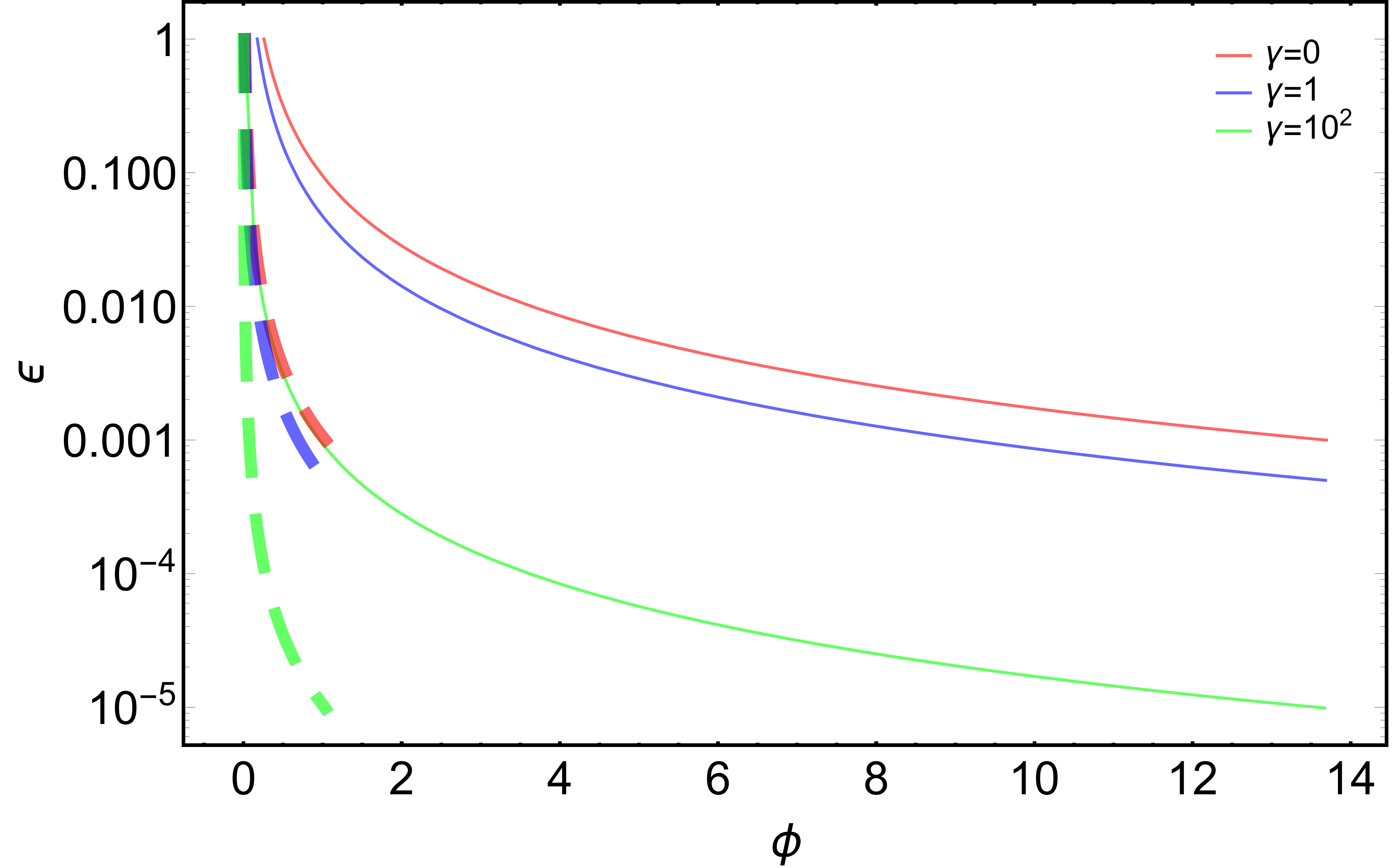}
	\caption{We plot the slow-roll parameter $\epsilon$ as a function of the inflaton field $\phi$ for several different values of the rainbow parameter $\gamma$ and  $\lambda$. The thick line corresponds to $\lambda/M_{pl}^{4-d}=10^{-12}$ and thin line $\lambda/M_{pl}^{4-d}=10^{-8}$. The relation \eqref{n_rela_1} gives us the allowed minimum value for $n$. In particular, for $N_{*}=50$ and $\gamma\in [0,\infty]$ we obtain $n>n_{min}\in [2.318,2.355]$. Then, as an example we take $N_{*}=50$ and $n=2.5$}.  
	\label{FIG2}
\end{figure}

The latest cosmological data from Planck satellite \cite{Planck:2018jri} give the following constraint for the scalar spectral index
\be
n_{s}=0.9649 \pm 0.0042,
\ee at $68 \%$ C.L., and combined with the BICEP2/Keck Array (BK14) \cite{BICEP2:2015nss,BICEP2:2018kqh}, for the tensor-to-scalar ratio yields 
\be
r<0.064,
\ee at $95 \%$ C.L. Therefore, by using the expressions \eqref{n_s_N} and \eqref{r_N}, and for $b=0$, we obtain that the amount of $e$-folds during inflation satisfies
\bea
50\leq N< 52.91,
\label{efolds1}
\eea with 
\be
d_{min}<d<d_{max},
\label{d_d}
\ee being
\begin{widetext}
\bea
&& d_{min}=\frac{n (N-32.36)}{n (N+64.72)-N-32.36} \Big[2+ \frac{1}{n (64.72 \gamma+(\gamma +1)N+64.72)-(\gamma+1) N-32.36 \gamma-32.86}\Big],
\eea
\bea
&& d_{max}=\frac{n N}{n (N+250)-N-125 } \Big[2+ \frac{1}{n (250 \gamma +\gamma N+N+250)-(1+\gamma) N-125.5-125 \gamma}\Big].
\eea
\end{widetext}

and
\bea
&& n>0.3942+\frac{5.599}{52.91-N}+\nonumber\\
&& \frac{1}{500-9.45 N+(500-9.45 N)\gamma}.
\label{n_rela_1}
\eea The latter relation gives a minimum value for $n$. For instance, for $N_{*}=50$ and $\gamma\in [0,\infty]$ one has $n>n_{min}\in [2.318,2.355]$. Thus, for $N_{*}=50$, $\gamma=1$ and $n=2.5$ we get $0.4320<d<0.4350$. Also, from Eq. \eqref{PsModel1}, and for $\xi M_{pl}^b=1$, we find  $2.505\times 10^9<\sigma<2.548 \times 10^9$ and $\sigma$ decreases as $d$ increases. In FIG. \ref{FIG1} we show the parameter $\sigma$ as a function of $d$, and for different values of $n$ and $\gamma$. The values of $\sigma$ increase very quickly  with $n$ and the rainbow effect is more important for smaller $n$. In FIG \ref{FIG2} we depict the behavior of the slow-roll parameter $\epsilon$ as a function of the inflaton field $\phi$. The rainbow parameter $\gamma$ has the effect of decreasing the parameter $\epsilon$.  

\begin{figure}[htbp]
	\centering
		\includegraphics[width=0.45\textwidth]{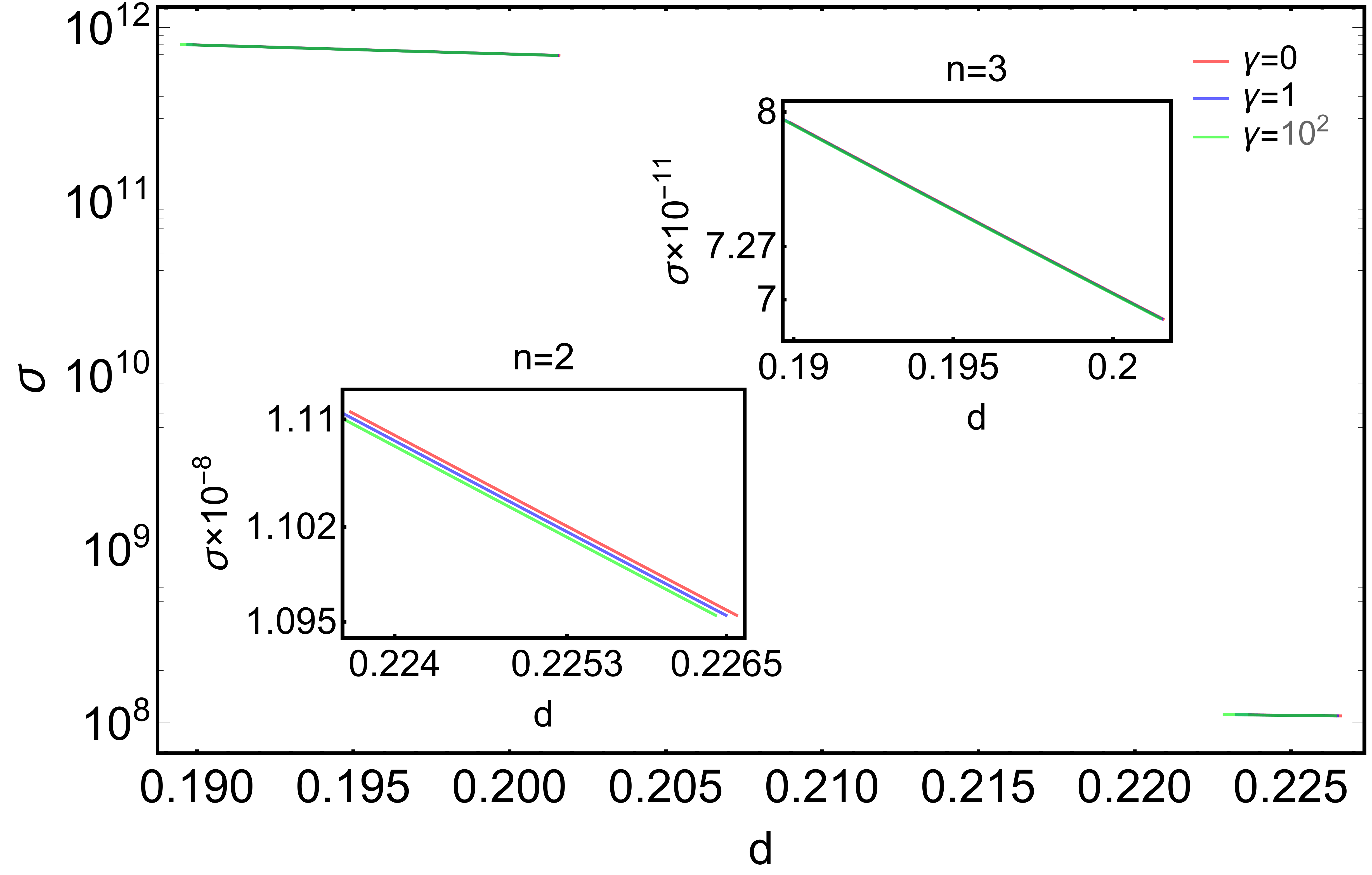}
	\caption{We depict the values of the parameter $\sigma$. We have assumed several different values of the rainbow parameter $\gamma$ and the power exponent $n$. Also, we used $N_{*}=40$ and the current observational value for the amplitude of primordial scalar perturbations $\mathcal{P}_{s}=2.141\times 10^{-9}$ \cite{Planck:2018jri}.}  
	\label{FIG3}
\end{figure}

\begin{figure}[htbp]
	\centering
		\includegraphics[width=0.45\textwidth]{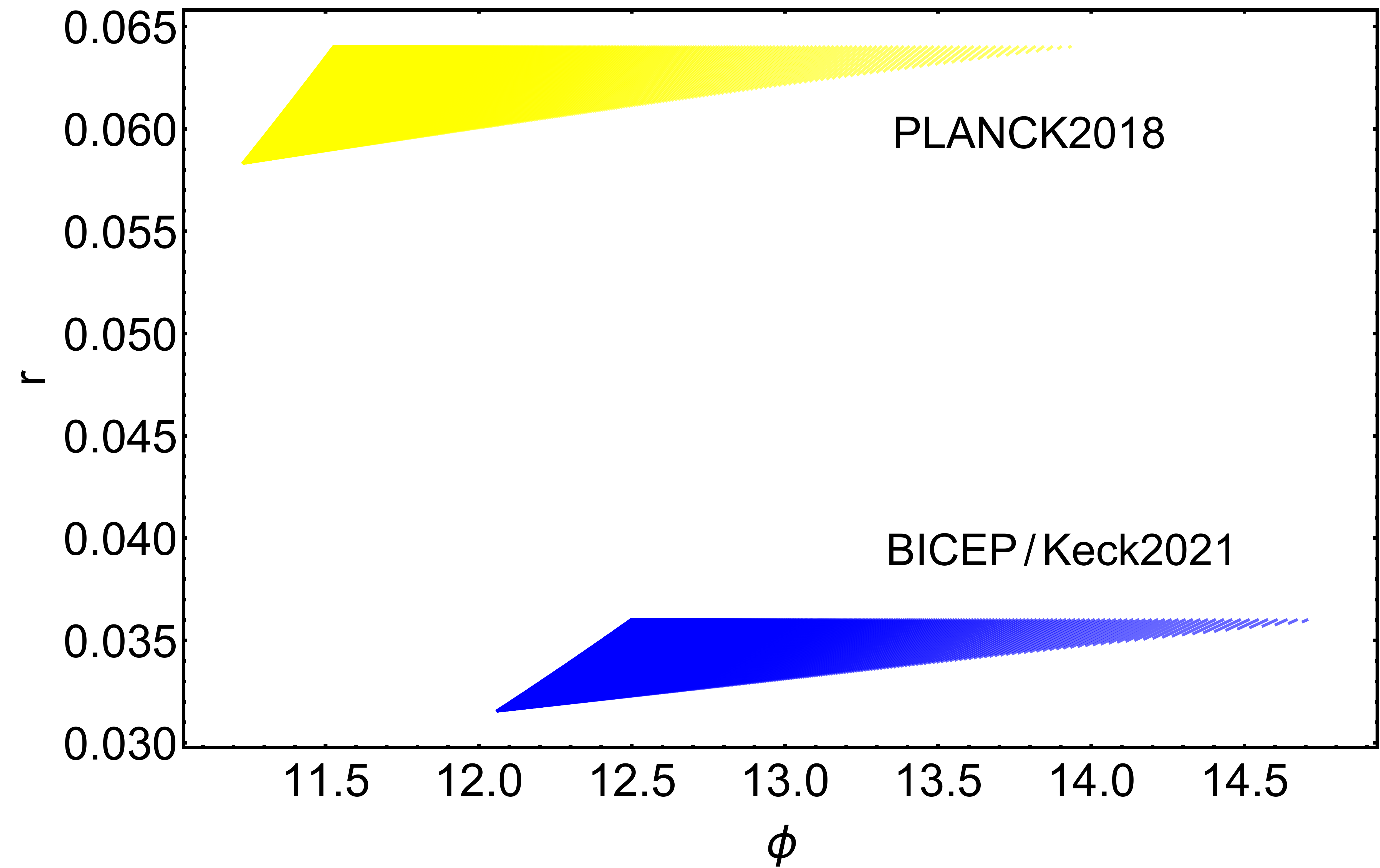}
	\caption{We depict the tensor-to-scalar ratio $r$ as a function of the inflaton $\phi$. The yellow region corresponds to $N_{*}=50$ and $\lambda/M_{pl}^{4-d}=1\times 10^{-8}$, and the blue one to $N_{*}=40$ and  $\lambda/M_{pl}^{4-d}=5\times 10^{-8}$. To confront with the predictions of the model we used both the PLANCK $2018$ data \cite{Planck:2018jri} and the recently released BICEP/Keck data \cite{BICEPKeck:2021gln}.}  
	\label{FIG4}
\end{figure}

New data from BICEP/ Keck XIII have recently been published \cite{BICEPKeck:2021gln} which put stronger constraints on the amplitude of primordial gravitational waves and then on the tensor-scalar ratio
\be
r<0.036,
\ee at $95 \%$ C.L. Thus, for $b=0$, we obtain
\be
40\leq  N < 41.41, 
\label{efolds2}
\ee with 
\small
\bea
&&d_{min}=\frac{n (1+\gamma) (2 N-64.72)}{(1+\gamma) \left[n (N+64.72)-N\right]-32.86-32.36 \gamma},\\
&& d_{max}=\frac{2 (1+\gamma) n N}{(1+\gamma) \left[n (N+444.4)-N\right]-222.2-222.7 \gamma},
\eea
\normalsize
and 
\bea
&& n > 0.4534+\frac{1.929}{41.41-N} +\nonumber\\
&& \frac{1}{888.9-21.47 N+\left(888.9-21.47 N\right)\gamma}.
\label{n_rela_2}
\eea
In this case, for $N_{*}=40$ and $\gamma \in [0,\infty]$ one obtains $n>n_{min} \in [1.824,1.857]$. So, for $N_{*}=40$, $\gamma=1$ and $n=2$ one finds $0.2233<d<0.2265$. Also, from Eq. \eqref{PsModel1}, and for $\xi M_{pl}^b=1$, we find  $1.095 \times 10^8<\sigma<1.112 \times 10^8$. In FIG. \ref{FIG3} we plot the parameter $\sigma$ as a function of $d$, and the other parameters $\gamma$ and $n$. By comparing with FIG. \ref{FIG1}, one can conclude that in this case smaller values of $d$ are required.
Similar behavior to that shown in FIG. \ref{FIG2} for the slow-roll parameter $\epsilon$ as a function of the inflaton field is found. Nevertheless, greater values of $\lambda$ are required in order to obtain a field displacement  equal to that of the previous case with $N_{*}=50$. 
Finally, in FIG. \ref{FIG4} we show the tensor-to-scalar ratio $r$ as a function of the inflaton field $\phi$ and use both the current PLANCK and BICEP/Keck data to confront the predictions of the model.

\section{Concluding Remarks}\label{conclusion_f}
In the present paper, we study inflation and the generation of primordial fluctuations in the context of $F(T)$ Gravity's Rainbow. 
After establishing the general setup to study slow-roll inflation in these theories, and by following Maldacena's method \cite{Maldacena:2002vr}, we calculated the second order action for scalar and tensor perturbations and their primordial power spectrum. By performing the latter step, we have assumed that the two energy-dependent rainbow functions $\tilde{f}$ and $\tilde{g}$ related to the energy and momentum generators for spacetime translations, respectively, are implicitly time dependent as the energy of the probe particles in an expanding universe can depend on the cosmic time \cite{Ling:2006az}. 

An important ingredient in our analysis is the violation of the local Lorentz symmetry in modified teleparallel gravity (MTG). Since the gravitational action of $F(T)$ gravity is not local Lorentz invariant \cite{Sotiriou:2010mv,Li:2010cg}, we need to take into account the corresponding additional six degrees of freedom in the analysis of the cosmological perturbations. These additional modes can be integrated out from the second order action for the perturbations, leading to the emergence of a new explicit mass term $\eta_{\mathcal{R}}= m^2/(3 H^2)$ ( $H$ the Hubble rate) in the scalar sector that describes the effects of the local Lorentz symmetry breaking in MTG \cite{Gonzalez-Espinoza:2020azh}. This explicit mass term $\eta_{\mathcal{R}}$ leads to a slight logarithmic time dependence and then to an additional scale dependence of the scalar power spectrum \cite{Gonzalez-Espinoza:2020azh,Riotto:2018pcx}. We found that in $F(T)$ Gravity's Rainbow this explicit mass term $\eta_{\mathcal{R}}$ may depend on the rainbow properties of the effective spacetime metric. Although $\eta_{\mathcal{R}}$ does not explicitly depend on the rainbow functions at first order of slow-roll approximation, it may depend implicitly through the solution $(H,\phi)$ for the background equations. This result also holds for the inflationary observables, namely, the scalar power spectrum $\mathcal{P}_{s}$, the scalar spectral index $n_{s}$ and the tensor-to-scalar ratio $r$. 

Since on a dynamical background, the rainbow functions $\tilde{f}$ and $\tilde{g}$ may depend on the cosmic time via the energy of the test particles \cite{Ling:2006az}, a natural assumption is to consider that they are functions of the Hubble rate $H$ \cite{Chatrabhuti:2015mws,Waeming:2020rir}. Thus, under the very reasonable assumption that the rainbow functions are power-law functions of $H$ \cite{Chatrabhuti:2015mws,Waeming:2020rir}, and for a power-law function $F(T)\sim T^n$ with a chaotic scalar potential $V\sim \phi^d$, we calculated the explicit mass term $\eta_{\mathcal{R}}$ and the inflationary observables $\mathcal{P}_{s}$, $n_{s}$ and $r$ as functions of the amount of $e$-folds during inflation, $N_{*}$. Although we found that in a general setup the function $\tilde{g}$ does not contribute to the dynamics of inflation up to first order in slow-roll approximation, we also found that the rainbow effect on $\eta_{\mathcal{R}}$ through the function $\tilde{f}$ becomes more important at intermediate times and at the end inflation by increasing the values of $\eta_{\mathcal{R}}$. However, since we are imposing the slow-roll initial conditions $\epsilon \ll 1$, $|\dot{\epsilon}|/H\epsilon \ll  1$, and then during inflation $\eta_{\mathcal{R}}\ll 1$, the curvature perturbation freezes on super-horizon scales and therefore we can evaluate the inflationary observables at the Hubble horizon crossing.

Thus, we used the Planck $2018$ \cite{Planck:2018jri} and BICEP/Keck $2021$ \cite{BICEPKeck:2021gln} data to constrain the free parameters of the model. Particularly, we found that the predictions of the model are consistent which both sets of data. In the case of the recently released BICEP/Keck $2021$ data, which puts considerably strengthened bounds on the tensor-to-scalar ratio $r$ \cite{Kallosh:2021mnu}, we found that the number of $e$-folds required by the model to be consistent with observations is smaller (with a flatter scalar potential) than in the case when applying the Planck $2018$ data. By assuming that reheating occurs in a very short time immediately  after a period of de Sitter inflation, the number of $e$-folds required to solve the flatness problem (and so the horizon problem) satisfies $\Delta{N}\gtrsim  64-\ln\left(10^{16}\text{GeV}/\rho_{\text{reh}}^{1/4}\right)$, where  $\rho_{\text{reh}}$ is the energy density at the end of reheating \cite{Maggiore:2018sht}. Then, for instantaneous reheating one obtains $\Delta{N}\gtrsim 64$, while for the lowest conceivable reheating temperature that is in agreement with the Big Bang Nucleosynthesis (BBN) requirements, $1~\text{MeV}$,  one gets $\Delta{N}\gtrsim 20$ \cite{Dai:2014jja,Munoz:2014eqa,Cook:2015vqa,Panotopoulos:2020qzi,Lopez:2021agu}. Therefore, the number of $e$-folds obtained in Eqs. \eqref{efolds1} and \eqref{efolds2} are viable to solve the early-universe problems of the standard hot Big-Bang cosmological model. However, let us remember that the precise value of the number of $e$-folds during inflation depends on the energy scale of inflation and also on the details of reheating after inflation \cite{Baumann:2018bq, Maggiore:2018sht} (see also Refs. \cite{Dai:2014jja,Munoz:2014eqa,Cook:2015vqa,Panotopoulos:2020qzi,Lopez:2021agu}). The study of reheating after inflation in the context of the present model lies beyond the scope of the present work and thus it is left for a separate project.

In the context of rainbow gravity and by appealing to the connection with the scenarios for the running of the spectral dimensions, in Ref. \cite{Amelino-Camelia:2013wha} (see also \cite{Amelino-Camelia:2013tla,Amelino-Camelia:2013gna}) the authors proposed the  modified dispersion relation (MDR) $\mathcal{E}^2=\mathbb{P}^2\left(1+(\tilde{\lambda} \mathbb{P})^{2 \tilde{\gamma}}\right)$, where $\tilde{\gamma}$ and $\tilde{\lambda}$ are constants. Particularly, by assuming Einstein Gravity and under some other considerations (e.g. higher order spatial derivatives are used) they showed this MDR with $\tilde{\gamma}=2$ leads to an exactly scale-invariant spectrum. Since the spectrum cannot be exactly scale invariant \cite{Planck:2018jri} they also studied small departures from the case $\tilde{\gamma}=2$ and proved that in the rainbow frame the effective equation of state is $w_{eff}=w-2 \tilde{\gamma}/3$.  Then an effective inflation is realized for a constant and non-inflationary equation of state $w<1$. On the other hand, for the MDR in Eq. \eqref{MDR}, with the rainbow functions \eqref{Rainbow_Function}, and in the presence of the matter field \eqref{L_matter}, we found $w_{eff}= w-\frac{\gamma  (w+1) (3 w+1)}{\gamma (3 w+1)-2}$, where  $w=-1-\frac{2 (p (c-d+2)-1)}{3 (1+p (d-c-2)+2 N)}$ is a function of the number of $e$-folds of inflation $N$. Thus, our results generalize those obtained in previous works to the case of a non-constant equation of state. However, an inflationary equation of state was used due to the matter content that was chosen. Indeed, for $N\gg 1$ one has $w\sim -1$ and then $w_{eff}\sim w \sim -1$. Interestingly, in our equations (e.g. Eq. \eqref{n_s_N}) the exactly scale invariant case corresponds to $\gamma=-1$ and then $w_{eff}=-1/3$. But this case was excluded in Eq. \eqref{Rainbow_Function} when the condition $H\gg M$ ($M$ is a constant with dimension of mass) was assumed, which is valid during inflation. For $n=1$ we recover the case of general relativity plus scalar field. We found that this case is disfavored by the current observational data even in the presence of the rainbow effects. For instance, from Eqs. \eqref{n_rela_1} and \eqref{n_rela_2} is deduced that the allowed minimum value for $n$ consistent with observations is greater than $n=1$. This latter result is in agreement with what has been obtained in the literature for the (pure) single field inflation model \cite{Planck:2018jri,BICEPKeck:2021gln}.


\begin{acknowledgments}
Y. L acknowledges acknowldeges Dirección de Investigación, Postgrado y Transferencia Tecnológica de la Universidad de Tarapacá for financial support through Proyecto UTA Mayor  No. 4740-20. G.O acknowldeges DI-VRIEA for financial support through Proyecto Postdoctorado $2020$ VRIEA-PUCV.

\end{acknowledgments}

\appendix

\bibliography{bio}

\begin{thebibliography}{99}
\expandafter\ifx\csname natexlab\endcsname\relax\def\natexlab#1{#1}\fi
\expandafter\ifx\csname bibnamefont\endcsname\relax
  \def\bibnamefont#1{#1}\fi
\expandafter\ifx\csname bibfnamefont\endcsname\relax
  \def\bibfnamefont#1{#1}\fi
\expandafter\ifx\csname citenamefont\endcsname\relax
  \def\citenamefont#1{#1}\fi
\expandafter\ifx\csname url\endcsname\relax
  \def\url#1{\texttt{#1}}\fi
\expandafter\ifx\csname urlprefix\endcsname\relax\def\urlprefix{URL }\fi
\providecommand{\bibinfo}[2]{#2}
\providecommand{\eprint}[2][]{\url{#2}}

\bibitem[{\citenamefont{Starobinsky}(1980)}]{Starobinsky}
\bibinfo{author}{\bibfnamefont{A.~A.} \bibnamefont{Starobinsky}},
  \bibinfo{journal}{Phys. Lett. B} \textbf{\bibinfo{volume}{91}},
  \bibinfo{pages}{99} (\bibinfo{year}{1980}).

\bibitem[{\citenamefont{Guth}(1981)}]{Guth:1980zm}
\bibinfo{author}{\bibfnamefont{A.~H.} \bibnamefont{Guth}},
  \bibinfo{journal}{Phys. Rev. D} \textbf{\bibinfo{volume}{23}},
  \bibinfo{pages}{347} (\bibinfo{year}{1981}).

\bibitem[{\citenamefont{Albrecht and Steinhardt}(1982)}]{Albrecht:1982wi}
\bibinfo{author}{\bibfnamefont{A.}~\bibnamefont{Albrecht}} \bibnamefont{and}
  \bibinfo{author}{\bibfnamefont{P.~J.} \bibnamefont{Steinhardt}},
  \bibinfo{journal}{Phys. Rev. Lett.} \textbf{\bibinfo{volume}{48}},
  \bibinfo{pages}{1220} (\bibinfo{year}{1982}).

\bibitem[{\citenamefont{Linde}(1982)}]{Linde:1981mu}
\bibinfo{author}{\bibfnamefont{A.~D.} \bibnamefont{Linde}},
  \bibinfo{journal}{Phys. Lett. B} \textbf{\bibinfo{volume}{108}},
  \bibinfo{pages}{389} (\bibinfo{year}{1982}).

\bibitem[{\citenamefont{Ade et~al.}(2016)}]{Planck:2015sxf}
\bibinfo{author}{\bibfnamefont{P.~A.~R.} \bibnamefont{Ade}}
  \bibnamefont{et~al.} (\bibinfo{collaboration}{Planck}),
  \bibinfo{journal}{Astron. Astrophys.} \textbf{\bibinfo{volume}{594}},
  \bibinfo{pages}{A20} (\bibinfo{year}{2016}), \eprint{1502.02114}.

\bibitem[{\citenamefont{Akrami et~al.}(2020)}]{Planck:2018jri}
\bibinfo{author}{\bibfnamefont{Y.}~\bibnamefont{Akrami}} \bibnamefont{et~al.}
  (\bibinfo{collaboration}{Planck}), \bibinfo{journal}{Astron. Astrophys.}
  \textbf{\bibinfo{volume}{641}}, \bibinfo{pages}{A10} (\bibinfo{year}{2020}),
  \eprint{1807.06211}.

\bibitem[{\citenamefont{Ade et~al.}(2015)}]{BICEP2:2015nss}
\bibinfo{author}{\bibfnamefont{P.~A.~R.} \bibnamefont{Ade}}
  \bibnamefont{et~al.} (\bibinfo{collaboration}{BICEP2, Planck}),
  \bibinfo{journal}{Phys. Rev. Lett.} \textbf{\bibinfo{volume}{114}},
  \bibinfo{pages}{101301} (\bibinfo{year}{2015}), \eprint{1502.00612}.

\bibitem[{\citenamefont{Ade et~al.}(2018)}]{BICEP2:2018kqh}
\bibinfo{author}{\bibfnamefont{P.~A.~R.} \bibnamefont{Ade}}
  \bibnamefont{et~al.} (\bibinfo{collaboration}{BICEP2, Keck Array}),
  \bibinfo{journal}{Phys. Rev. Lett.} \textbf{\bibinfo{volume}{121}},
  \bibinfo{pages}{221301} (\bibinfo{year}{2018}), \eprint{1810.05216}.

\bibitem[{\citenamefont{Senatore}(2017)}]{Senatore:2016aui}
\bibinfo{author}{\bibfnamefont{L.}~\bibnamefont{Senatore}}, in
  \emph{\bibinfo{booktitle}{{Proceedings, Theoretical Advanced Study Institute
  in Elementary Particle Physics: New Frontiers in Fields and Strings (TASI
  2015)}: {Boulder, CO, USA, June 1-26, 2015}}} (\bibinfo{publisher}{WSP},
  \bibinfo{address}{Singapore}, \bibinfo{year}{2017}), pp.
  \bibinfo{pages}{447--543}.

\bibitem[{\citenamefont{Riotto}(2003)}]{Riotto:2018pcx}
\bibinfo{author}{\bibfnamefont{A.}~\bibnamefont{Riotto}},
  \bibinfo{journal}{ICTP Lect. Notes Ser.} \textbf{\bibinfo{volume}{14}},
  \bibinfo{pages}{317} (\bibinfo{year}{2003}), \eprint{hep-ph/0210162}.

\bibitem[{\citenamefont{Baumann}(2018)}]{Baumann:2018bq}
\bibinfo{author}{\bibfnamefont{D.}~\bibnamefont{Baumann}}, in
  \emph{\bibinfo{booktitle}{{Proceedings of Theoretical Advanced Study
  Institute Summer School 2017, "Physics at the Fundamental Frontier"}
  {\textemdash} PoS(TASI2017)}} (\bibinfo{year}{2018}), vol.
  \bibinfo{volume}{305}, p. \bibinfo{pages}{009}.

\bibitem[{\citenamefont{Baumann and McAllister}(2015)}]{Baumann:2014nda}
\bibinfo{author}{\bibfnamefont{D.}~\bibnamefont{Baumann}} \bibnamefont{and}
  \bibinfo{author}{\bibfnamefont{L.}~\bibnamefont{McAllister}},
  \emph{\bibinfo{title}{{Inflation and String Theory}}}, Cambridge Monographs
  on Mathematical Physics (\bibinfo{publisher}{Cambridge University Press},
  \bibinfo{year}{2015}), ISBN \bibinfo{isbn}{9781107089693, 9781316237182},
  \eprint{1404.2601}.

\bibitem[{\citenamefont{Weinberg}(2008)}]{Weinberg:2008zzc}
\bibinfo{author}{\bibfnamefont{S.}~\bibnamefont{Weinberg}},
  \emph{\bibinfo{title}{{Cosmology}}} (\bibinfo{publisher}{Oxford Univ. Press},
  \bibinfo{year}{2008}), ISBN \bibinfo{isbn}{978-0-19-852682-7}.

\bibitem[{\citenamefont{Maggiore}(2018)}]{Maggiore:2018sht}
\bibinfo{author}{\bibfnamefont{M.}~\bibnamefont{Maggiore}},
  \emph{\bibinfo{title}{{Gravitational Waves. Vol. 2: Astrophysics and
  Cosmology}}} (\bibinfo{publisher}{Oxford University Press},
  \bibinfo{year}{2018}), ISBN \bibinfo{isbn}{978-0-19-857089-9}.

\bibitem[{\citenamefont{Ade et~al.}(2021)}]{BICEPKeck:2021gln}
\bibinfo{author}{\bibfnamefont{P.~A.~R.} \bibnamefont{Ade}}
  \bibnamefont{et~al.} (\bibinfo{collaboration}{BICEP/Keck}),
  \bibinfo{journal}{Phys. Rev. Lett.} \textbf{\bibinfo{volume}{127}},
  \bibinfo{pages}{151301} (\bibinfo{year}{2021}), \eprint{2110.00483}.

\bibitem[{\citenamefont{Weinberg}(1972)}]{Weinberg:1972kfs}
\bibinfo{author}{\bibfnamefont{S.}~\bibnamefont{Weinberg}},
  \emph{\bibinfo{title}{{Gravitation and Cosmology}: {Principles and
  Applications of the General Theory of Relativity}}} (\bibinfo{publisher}{John
  Wiley and Sons}, \bibinfo{address}{New York}, \bibinfo{year}{1972}), ISBN
  \bibinfo{isbn}{978-0-471-92567-5, 978-0-471-92567-5}.

\bibitem[{\citenamefont{Wald}(1984)}]{Wald:1984rg}
\bibinfo{author}{\bibfnamefont{R.~M.} \bibnamefont{Wald}},
  \emph{\bibinfo{title}{{General Relativity}}} (\bibinfo{publisher}{Chicago
  Univ. Pr.}, \bibinfo{address}{Chicago, USA}, \bibinfo{year}{1984}).

\bibitem[{\citenamefont{Einstein}(1928)}]{Einstein}
\bibinfo{author}{\bibfnamefont{A.}~\bibnamefont{Einstein}},
  \bibinfo{journal}{Sitzungsber. Preuss. Akad. Wiss. Phys. Math. Kl.}
  \textbf{\bibinfo{volume}{17}}, \bibinfo{pages}{217} (\bibinfo{year}{1928}).

\bibitem[{\citenamefont{Unzicker and Case}(2005)}]{TranslationEinstein}
\bibinfo{author}{\bibfnamefont{A.}~\bibnamefont{Unzicker}} \bibnamefont{and}
  \bibinfo{author}{\bibfnamefont{T.}~\bibnamefont{Case}},
  \bibinfo{journal}{arXiv:physics/0503046}  (\bibinfo{year}{2005}).

\bibitem[{\citenamefont{Einstein}(1930{\natexlab{a}})}]{Early-papers1}
\bibinfo{author}{\bibfnamefont{A.}~\bibnamefont{Einstein}},
  \bibinfo{journal}{Math. Ann.} \textbf{\bibinfo{volume}{102}},
  \bibinfo{pages}{685} (\bibinfo{year}{1930}{\natexlab{a}}).

\bibitem[{\citenamefont{Einstein}(1930{\natexlab{b}})}]{Early-papers2}
\bibinfo{author}{\bibfnamefont{A.}~\bibnamefont{Einstein}},
  \bibinfo{journal}{Sitzungsber. Preuss. Akad. Wiss. Phys. Math. Kl.}
  \textbf{\bibinfo{volume}{24}}, \bibinfo{pages}{401}
  (\bibinfo{year}{1930}{\natexlab{b}}).

\bibitem[{\citenamefont{Pellegrini and Plebanski}(1963)}]{Early-papers3}
\bibinfo{author}{\bibfnamefont{C.}~\bibnamefont{Pellegrini}} \bibnamefont{and}
  \bibinfo{author}{\bibfnamefont{J.}~\bibnamefont{Plebanski}},
  \bibinfo{journal}{Math.-Fys. Skr. Dan. Vid. Selskab}
  \textbf{\bibinfo{volume}{2}}, \bibinfo{pages}{4} (\bibinfo{year}{1963}).

\bibitem[{\citenamefont{M{\o}ller}(1978)}]{Early-papers4}
\bibinfo{author}{\bibfnamefont{C.}~\bibnamefont{M{\o}ller}},
  \bibinfo{journal}{K. Dan.Vidensk. Selsk. Mat. Fys. Skr.}
  \textbf{\bibinfo{volume}{89}}, \bibinfo{pages}{13} (\bibinfo{year}{1978}).

\bibitem[{\citenamefont{Hayashi and Nakano}(1967)}]{Early-papers5}
\bibinfo{author}{\bibfnamefont{K.}~\bibnamefont{Hayashi}} \bibnamefont{and}
  \bibinfo{author}{\bibfnamefont{T.}~\bibnamefont{Nakano}},
  \bibinfo{journal}{Progress of Theoretical Physics}
  \textbf{\bibinfo{volume}{38}}, \bibinfo{pages}{491} (\bibinfo{year}{1967}).

\bibitem[{\citenamefont{Hayashi and Shirafuji}(1979)}]{Early-papers6}
\bibinfo{author}{\bibfnamefont{K.}~\bibnamefont{Hayashi}} \bibnamefont{and}
  \bibinfo{author}{\bibfnamefont{T.}~\bibnamefont{Shirafuji}},
  \bibinfo{journal}{Phys. Rev. D} \textbf{\bibinfo{volume}{19}},
  \bibinfo{pages}{3524} (\bibinfo{year}{1979}).

\bibitem[{\citenamefont{Pereira}(2014)}]{JGPereira2}
\bibinfo{author}{\bibfnamefont{J.~G.} \bibnamefont{Pereira}}, in
  \emph{\bibinfo{booktitle}{Handbook of Spacetime}}, edited by
  \bibinfo{editor}{\bibfnamefont{A.}~\bibnamefont{Ashtekar}} \bibnamefont{and}
  \bibinfo{editor}{\bibfnamefont{V.}~\bibnamefont{Petkov}}
  (\bibinfo{publisher}{Springer}, \bibinfo{year}{2014}), pp.
  \bibinfo{pages}{197--212}, \eprint{1302.6983}.

\bibitem[{\citenamefont{de~Andrade et~al.}(2000)\citenamefont{de~Andrade,
  Guillen, and Pereira}}]{AndradeGuillenPereira-00}
\bibinfo{author}{\bibfnamefont{V.~C.} \bibnamefont{de~Andrade}},
  \bibinfo{author}{\bibfnamefont{L.~C.~T.} \bibnamefont{Guillen}},
  \bibnamefont{and} \bibinfo{author}{\bibfnamefont{J.~G.}
  \bibnamefont{Pereira}}, \bibinfo{journal}{Phys. Rev. Lett.}
  \textbf{\bibinfo{volume}{84}}, \bibinfo{pages}{4533} (\bibinfo{year}{2000}),
  \eprint{gr-qc/0003100}.

\bibitem[{\citenamefont{Arcos and Pereira}(2004)}]{Arcos:2005ec}
\bibinfo{author}{\bibfnamefont{H.~I.} \bibnamefont{Arcos}} \bibnamefont{and}
  \bibinfo{author}{\bibfnamefont{J.~G.} \bibnamefont{Pereira}},
  \bibinfo{journal}{Int. J. Mod. Phys. D} \textbf{\bibinfo{volume}{13}},
  \bibinfo{pages}{2193} (\bibinfo{year}{2004}), \eprint{gr-qc/0501017}.

\bibitem[{\citenamefont{Pereira and Obukhov}(2019)}]{Pereira:2019woq}
\bibinfo{author}{\bibfnamefont{J.~G.} \bibnamefont{Pereira}} \bibnamefont{and}
  \bibinfo{author}{\bibfnamefont{Y.~N.} \bibnamefont{Obukhov}},
  \bibinfo{journal}{Universe} \textbf{\bibinfo{volume}{5}},
  \bibinfo{pages}{139} (\bibinfo{year}{2019}), \eprint{1906.06287}.

\bibitem[{\citenamefont{Aldrovandi and
  Pereira}(2012)}]{Aldrovandi-Pereira-book}
\bibinfo{author}{\bibfnamefont{R.}~\bibnamefont{Aldrovandi}} \bibnamefont{and}
  \bibinfo{author}{\bibfnamefont{J.~G.} \bibnamefont{Pereira}},
  \emph{\bibinfo{title}{Teleparallel gravity: an introduction}}, vol.
  \bibinfo{volume}{173} (\bibinfo{publisher}{Springer Science \& Business
  Media}, \bibinfo{year}{2012}).

\bibitem[{\citenamefont{Cai et~al.}(2016)\citenamefont{Cai, Capozziello,
  De~Laurentis, and Saridakis}}]{Cai:2015emx}
\bibinfo{author}{\bibfnamefont{Y.-F.} \bibnamefont{Cai}},
  \bibinfo{author}{\bibfnamefont{S.}~\bibnamefont{Capozziello}},
  \bibinfo{author}{\bibfnamefont{M.}~\bibnamefont{De~Laurentis}},
  \bibnamefont{and} \bibinfo{author}{\bibfnamefont{E.~N.}
  \bibnamefont{Saridakis}}, \bibinfo{journal}{Rept. Prog. Phys.}
  \textbf{\bibinfo{volume}{79}}, \bibinfo{pages}{106901}
  (\bibinfo{year}{2016}), \eprint{1511.07586}.

\bibitem[{\citenamefont{Bahamonde et~al.}(2021)\citenamefont{Bahamonde,
  Dialektopoulos, Escamilla-Rivera, Farrugia, Gakis, Hendry, Hohmann, Said,
  Mifsud, and Di~Valentino}}]{Bahamonde:2021gfp}
\bibinfo{author}{\bibfnamefont{S.}~\bibnamefont{Bahamonde}},
  \bibinfo{author}{\bibfnamefont{K.~F.} \bibnamefont{Dialektopoulos}},
  \bibinfo{author}{\bibfnamefont{C.}~\bibnamefont{Escamilla-Rivera}},
  \bibinfo{author}{\bibfnamefont{G.}~\bibnamefont{Farrugia}},
  \bibinfo{author}{\bibfnamefont{V.}~\bibnamefont{Gakis}},
  \bibinfo{author}{\bibfnamefont{M.}~\bibnamefont{Hendry}},
  \bibinfo{author}{\bibfnamefont{M.}~\bibnamefont{Hohmann}},
  \bibinfo{author}{\bibfnamefont{J.~L.} \bibnamefont{Said}},
  \bibinfo{author}{\bibfnamefont{J.}~\bibnamefont{Mifsud}}, \bibnamefont{and}
  \bibinfo{author}{\bibfnamefont{E.}~\bibnamefont{Di~Valentino}}
  (\bibinfo{year}{2021}), \eprint{2106.13793}.

\bibitem[{\citenamefont{Bengochea and Ferraro}(2009)}]{Bengochea:2008gz}
\bibinfo{author}{\bibfnamefont{G.~R.} \bibnamefont{Bengochea}}
  \bibnamefont{and} \bibinfo{author}{\bibfnamefont{R.}~\bibnamefont{Ferraro}},
  \bibinfo{journal}{Phys. Rev.} \textbf{\bibinfo{volume}{D79}},
  \bibinfo{pages}{124019} (\bibinfo{year}{2009}), \eprint{0812.1205}.

\bibitem[{\citenamefont{Linder}(2010)}]{Linder:2010py}
\bibinfo{author}{\bibfnamefont{E.~V.} \bibnamefont{Linder}},
  \bibinfo{journal}{Phys. Rev.} \textbf{\bibinfo{volume}{D81}},
  \bibinfo{pages}{127301} (\bibinfo{year}{2010}), \eprint{1005.3039}.

\bibitem[{\citenamefont{Li et~al.}(2011{\natexlab{a}})\citenamefont{Li,
  Sotiriou, and Barrow}}]{Li:2011wu}
\bibinfo{author}{\bibfnamefont{B.}~\bibnamefont{Li}},
  \bibinfo{author}{\bibfnamefont{T.~P.} \bibnamefont{Sotiriou}},
  \bibnamefont{and} \bibinfo{author}{\bibfnamefont{J.~D.}
  \bibnamefont{Barrow}}, \bibinfo{journal}{Phys. Rev. D}
  \textbf{\bibinfo{volume}{83}}, \bibinfo{pages}{104017}
  (\bibinfo{year}{2011}{\natexlab{a}}), \eprint{1103.2786}.

\bibitem[{\citenamefont{Clifton et~al.}(2012)\citenamefont{Clifton, Ferreira,
  Padilla, and Skordis}}]{Clifton:2011jh}
\bibinfo{author}{\bibfnamefont{T.}~\bibnamefont{Clifton}},
  \bibinfo{author}{\bibfnamefont{P.~G.} \bibnamefont{Ferreira}},
  \bibinfo{author}{\bibfnamefont{A.}~\bibnamefont{Padilla}}, \bibnamefont{and}
  \bibinfo{author}{\bibfnamefont{C.}~\bibnamefont{Skordis}},
  \bibinfo{journal}{Phys. Rept.} \textbf{\bibinfo{volume}{513}},
  \bibinfo{pages}{1} (\bibinfo{year}{2012}), \eprint{1106.2476}.

\bibitem[{\citenamefont{Capozziello and
  De~Laurentis}(2011)}]{Capozziello:2011et}
\bibinfo{author}{\bibfnamefont{S.}~\bibnamefont{Capozziello}} \bibnamefont{and}
  \bibinfo{author}{\bibfnamefont{M.}~\bibnamefont{De~Laurentis}},
  \bibinfo{journal}{Phys. Rept.} \textbf{\bibinfo{volume}{509}},
  \bibinfo{pages}{167} (\bibinfo{year}{2011}), \eprint{1108.6266}.

\bibitem[{\citenamefont{De~Felice and Tsujikawa}(2010)}]{DeFelice:2010aj}
\bibinfo{author}{\bibfnamefont{A.}~\bibnamefont{De~Felice}} \bibnamefont{and}
  \bibinfo{author}{\bibfnamefont{S.}~\bibnamefont{Tsujikawa}},
  \bibinfo{journal}{Living Rev. Rel.} \textbf{\bibinfo{volume}{13}},
  \bibinfo{pages}{3} (\bibinfo{year}{2010}), \eprint{1002.4928}.

\bibitem[{\citenamefont{Nojiri and Odintsov}(2011)}]{Nojiri:2010wj}
\bibinfo{author}{\bibfnamefont{S.}~\bibnamefont{Nojiri}} \bibnamefont{and}
  \bibinfo{author}{\bibfnamefont{S.~D.} \bibnamefont{Odintsov}},
  \bibinfo{journal}{Phys. Rept.} \textbf{\bibinfo{volume}{505}},
  \bibinfo{pages}{59} (\bibinfo{year}{2011}), \eprint{1011.0544}.

\bibitem[{\citenamefont{Nojiri and Odintsov}(2006)}]{Nojiri:2006ri}
\bibinfo{author}{\bibfnamefont{S.}~\bibnamefont{Nojiri}} \bibnamefont{and}
  \bibinfo{author}{\bibfnamefont{S.~D.} \bibnamefont{Odintsov}},
  \bibinfo{journal}{eConf} \textbf{\bibinfo{volume}{C0602061}},
  \bibinfo{pages}{06} (\bibinfo{year}{2006}), \eprint{hep-th/0601213}.

\bibitem[{\citenamefont{Yerzhanov et~al.}(2010)\citenamefont{Yerzhanov,
  Myrzakul, Kulnazarov, and Myrzakulov}}]{Yerzhanov:2010vu}
\bibinfo{author}{\bibfnamefont{K.~K.} \bibnamefont{Yerzhanov}},
  \bibinfo{author}{\bibfnamefont{S.~R.} \bibnamefont{Myrzakul}},
  \bibinfo{author}{\bibfnamefont{I.~I.} \bibnamefont{Kulnazarov}},
  \bibnamefont{and}
  \bibinfo{author}{\bibfnamefont{R.}~\bibnamefont{Myrzakulov}},
  \bibinfo{journal}{arXiv:1006.3879}  (\bibinfo{year}{2010}),
  \eprint{1006.3879}.

\bibitem[{\citenamefont{Chakrabarti et~al.}(2017)\citenamefont{Chakrabarti,
  Said, and Farrugia}}]{Chakrabarti:2017moe}
\bibinfo{author}{\bibfnamefont{S.}~\bibnamefont{Chakrabarti}},
  \bibinfo{author}{\bibfnamefont{J.~L.} \bibnamefont{Said}}, \bibnamefont{and}
  \bibinfo{author}{\bibfnamefont{G.}~\bibnamefont{Farrugia}},
  \bibinfo{journal}{Eur. Phys. J. C} \textbf{\bibinfo{volume}{77}},
  \bibinfo{pages}{815} (\bibinfo{year}{2017}), \eprint{1711.04423}.

\bibitem[{\citenamefont{Hohmann et~al.}(2018)\citenamefont{Hohmann, J\"arv, and
  Ualikhanova}}]{Hohmann:2018rwf}
\bibinfo{author}{\bibfnamefont{M.}~\bibnamefont{Hohmann}},
  \bibinfo{author}{\bibfnamefont{L.}~\bibnamefont{J\"arv}}, \bibnamefont{and}
  \bibinfo{author}{\bibfnamefont{U.}~\bibnamefont{Ualikhanova}},
  \bibinfo{journal}{Phys. Rev. D} \textbf{\bibinfo{volume}{97}},
  \bibinfo{pages}{104011} (\bibinfo{year}{2018}), \eprint{1801.05786}.

\bibitem[{\citenamefont{Gonzalez-Espinoza and
  Otalora}(2020)}]{Gonzalez-Espinoza:2020azh}
\bibinfo{author}{\bibfnamefont{M.}~\bibnamefont{Gonzalez-Espinoza}}
  \bibnamefont{and} \bibinfo{author}{\bibfnamefont{G.}~\bibnamefont{Otalora}},
  \bibinfo{journal}{Phys. Lett. B} \textbf{\bibinfo{volume}{809}},
  \bibinfo{pages}{135696} (\bibinfo{year}{2020}), \eprint{2005.03753}.

\bibitem[{\citenamefont{Rezazadeh et~al.}(2016)\citenamefont{Rezazadeh,
  Abdolmaleki, and Karami}}]{Rezazadeh:2015dza}
\bibinfo{author}{\bibfnamefont{K.}~\bibnamefont{Rezazadeh}},
  \bibinfo{author}{\bibfnamefont{A.}~\bibnamefont{Abdolmaleki}},
  \bibnamefont{and} \bibinfo{author}{\bibfnamefont{K.}~\bibnamefont{Karami}},
  \bibinfo{journal}{JHEP} \textbf{\bibinfo{volume}{01}}, \bibinfo{pages}{131}
  (\bibinfo{year}{2016}), \eprint{1509.08769}.

\bibitem[{\citenamefont{Goodarzi and Mohseni~Sadjadi}(2019)}]{Goodarzi:2018feh}
\bibinfo{author}{\bibfnamefont{P.}~\bibnamefont{Goodarzi}} \bibnamefont{and}
  \bibinfo{author}{\bibfnamefont{H.}~\bibnamefont{Mohseni~Sadjadi}},
  \bibinfo{journal}{Eur. Phys. J. C} \textbf{\bibinfo{volume}{79}},
  \bibinfo{pages}{193} (\bibinfo{year}{2019}), \eprint{1808.01225}.

\bibitem[{\citenamefont{Bamba et~al.}(2016)\citenamefont{Bamba, Nashed,
  El~Hanafy, and Ibraheem}}]{Bamba:2016gbu}
\bibinfo{author}{\bibfnamefont{K.}~\bibnamefont{Bamba}},
  \bibinfo{author}{\bibfnamefont{G.~G.~L.} \bibnamefont{Nashed}},
  \bibinfo{author}{\bibfnamefont{W.}~\bibnamefont{El~Hanafy}},
  \bibnamefont{and} \bibinfo{author}{\bibfnamefont{S.~K.}
  \bibnamefont{Ibraheem}}, \bibinfo{journal}{Phys. Rev. D}
  \textbf{\bibinfo{volume}{94}}, \bibinfo{pages}{083513}
  (\bibinfo{year}{2016}), \eprint{1604.07604}.

\bibitem[{\citenamefont{Geng et~al.}(2011)\citenamefont{Geng, Lee, Saridakis,
  and Wu}}]{Geng:2011aj}
\bibinfo{author}{\bibfnamefont{C.-Q.} \bibnamefont{Geng}},
  \bibinfo{author}{\bibfnamefont{C.-C.} \bibnamefont{Lee}},
  \bibinfo{author}{\bibfnamefont{E.~N.} \bibnamefont{Saridakis}},
  \bibnamefont{and} \bibinfo{author}{\bibfnamefont{Y.-P.} \bibnamefont{Wu}},
  \bibinfo{journal}{Phys. Lett. B} \textbf{\bibinfo{volume}{704}},
  \bibinfo{pages}{384} (\bibinfo{year}{2011}), \eprint{1109.1092}.

\bibitem[{\citenamefont{Geng et~al.}(2012)\citenamefont{Geng, Lee, and
  Saridakis}}]{Geng:2011ka}
\bibinfo{author}{\bibfnamefont{C.-Q.} \bibnamefont{Geng}},
  \bibinfo{author}{\bibfnamefont{C.-C.} \bibnamefont{Lee}}, \bibnamefont{and}
  \bibinfo{author}{\bibfnamefont{E.~N.} \bibnamefont{Saridakis}},
  \bibinfo{journal}{JCAP} \textbf{\bibinfo{volume}{1201}}, \bibinfo{pages}{002}
  (\bibinfo{year}{2012}), \eprint{1110.0913}.

\bibitem[{\citenamefont{Xu et~al.}(2012)\citenamefont{Xu, Saridakis, and
  Leon}}]{Xu:2012jf}
\bibinfo{author}{\bibfnamefont{C.}~\bibnamefont{Xu}},
  \bibinfo{author}{\bibfnamefont{E.~N.} \bibnamefont{Saridakis}},
  \bibnamefont{and} \bibinfo{author}{\bibfnamefont{G.}~\bibnamefont{Leon}},
  \bibinfo{journal}{JCAP} \textbf{\bibinfo{volume}{1207}}, \bibinfo{pages}{005}
  (\bibinfo{year}{2012}), \eprint{1202.3781}.

\bibitem[{\citenamefont{Wei}(2012)}]{Wei:2011yr}
\bibinfo{author}{\bibfnamefont{H.}~\bibnamefont{Wei}}, \bibinfo{journal}{Phys.
  Lett. B} \textbf{\bibinfo{volume}{712}}, \bibinfo{pages}{430}
  (\bibinfo{year}{2012}), \eprint{1109.6107}.

\bibitem[{\citenamefont{Otalora}(2013{\natexlab{a}})}]{Otalora:2013tba}
\bibinfo{author}{\bibfnamefont{G.}~\bibnamefont{Otalora}},
  \bibinfo{journal}{JCAP} \textbf{\bibinfo{volume}{1307}}, \bibinfo{pages}{044}
  (\bibinfo{year}{2013}{\natexlab{a}}), \eprint{1305.0474}.

\bibitem[{\citenamefont{Otalora}(2013{\natexlab{b}})}]{Otalora:2013dsa}
\bibinfo{author}{\bibfnamefont{G.}~\bibnamefont{Otalora}},
  \bibinfo{journal}{Phys. Rev. D} \textbf{\bibinfo{volume}{88}},
  \bibinfo{pages}{063505} (\bibinfo{year}{2013}{\natexlab{b}}),
  \eprint{1305.5896}.

\bibitem[{\citenamefont{Otalora}(2015)}]{Otalora:2014aoa}
\bibinfo{author}{\bibfnamefont{G.}~\bibnamefont{Otalora}},
  \bibinfo{journal}{Int. J. Mod. Phys. D} \textbf{\bibinfo{volume}{25}},
  \bibinfo{pages}{1650025} (\bibinfo{year}{2015}), \eprint{1402.2256}.

\bibitem[{\citenamefont{Otalora Pati\~no}(2014)}]{OtaloraPatino:2014aru}
\bibinfo{author}{\bibfnamefont{G.}~\bibnamefont{Otalora Pati\~no}}, Ph.D.
  thesis, \bibinfo{school}{Sao Paulo, IFT} (\bibinfo{year}{2014}).

\bibitem[{\citenamefont{Skugoreva et~al.}(2015)\citenamefont{Skugoreva,
  Saridakis, and Toporensky}}]{Skugoreva:2014ena}
\bibinfo{author}{\bibfnamefont{M.~A.} \bibnamefont{Skugoreva}},
  \bibinfo{author}{\bibfnamefont{E.~N.} \bibnamefont{Saridakis}},
  \bibnamefont{and} \bibinfo{author}{\bibfnamefont{A.~V.}
  \bibnamefont{Toporensky}}, \bibinfo{journal}{Phys. Rev. D}
  \textbf{\bibinfo{volume}{91}}, \bibinfo{pages}{044023}
  (\bibinfo{year}{2015}), \eprint{1412.1502}.

\bibitem[{\citenamefont{Jarv and Toporensky}(2016)}]{Jarv:2015odu}
\bibinfo{author}{\bibfnamefont{L.}~\bibnamefont{Jarv}} \bibnamefont{and}
  \bibinfo{author}{\bibfnamefont{A.}~\bibnamefont{Toporensky}},
  \bibinfo{journal}{Phys. Rev. D} \textbf{\bibinfo{volume}{93}},
  \bibinfo{pages}{024051} (\bibinfo{year}{2016}), \eprint{1511.03933}.

\bibitem[{\citenamefont{Gonzalez-Espinoza
  et~al.}(2019)\citenamefont{Gonzalez-Espinoza, Otalora, Videla, and
  Saavedra}}]{Gonzalez-Espinoza:2019ajd}
\bibinfo{author}{\bibfnamefont{M.}~\bibnamefont{Gonzalez-Espinoza}},
  \bibinfo{author}{\bibfnamefont{G.}~\bibnamefont{Otalora}},
  \bibinfo{author}{\bibfnamefont{N.}~\bibnamefont{Videla}}, \bibnamefont{and}
  \bibinfo{author}{\bibfnamefont{J.}~\bibnamefont{Saavedra}},
  \bibinfo{journal}{JCAP} \textbf{\bibinfo{volume}{08}}, \bibinfo{pages}{029}
  (\bibinfo{year}{2019}), \eprint{1904.08068}.

\bibitem[{\citenamefont{Gonzalez-Espinoza and
  Otalora}(2021)}]{Gonzalez-Espinoza:2020jss}
\bibinfo{author}{\bibfnamefont{M.}~\bibnamefont{Gonzalez-Espinoza}}
  \bibnamefont{and} \bibinfo{author}{\bibfnamefont{G.}~\bibnamefont{Otalora}},
  \bibinfo{journal}{Eur. Phys. J. C} \textbf{\bibinfo{volume}{81}},
  \bibinfo{pages}{480} (\bibinfo{year}{2021}), \eprint{2011.08377}.

\bibitem[{\citenamefont{Gonzalez-Espinoza
  et~al.}(2021{\natexlab{a}})\citenamefont{Gonzalez-Espinoza, Herrera, Otalora,
  and Saavedra}}]{Gonzalez-Espinoza:2021qnv}
\bibinfo{author}{\bibfnamefont{M.}~\bibnamefont{Gonzalez-Espinoza}},
  \bibinfo{author}{\bibfnamefont{R.}~\bibnamefont{Herrera}},
  \bibinfo{author}{\bibfnamefont{G.}~\bibnamefont{Otalora}}, \bibnamefont{and}
  \bibinfo{author}{\bibfnamefont{J.}~\bibnamefont{Saavedra}},
  \bibinfo{journal}{Eur. Phys. J. C} \textbf{\bibinfo{volume}{81}},
  \bibinfo{pages}{731} (\bibinfo{year}{2021}{\natexlab{a}}),
  \eprint{2106.06145}.

\bibitem[{\citenamefont{Gonzalez-Espinoza
  et~al.}(2021{\natexlab{b}})\citenamefont{Gonzalez-Espinoza, Otalora, and
  Saavedra}}]{Gonzalez-Espinoza:2021mwr}
\bibinfo{author}{\bibfnamefont{M.}~\bibnamefont{Gonzalez-Espinoza}},
  \bibinfo{author}{\bibfnamefont{G.}~\bibnamefont{Otalora}}, \bibnamefont{and}
  \bibinfo{author}{\bibfnamefont{J.}~\bibnamefont{Saavedra}},
  \bibinfo{journal}{JCAP} \textbf{\bibinfo{volume}{10}}, \bibinfo{pages}{007}
  (\bibinfo{year}{2021}{\natexlab{b}}), \eprint{2101.09123}.

\bibitem[{\citenamefont{Magueijo and Smolin}(2004)}]{Magueijo:2002xx}
\bibinfo{author}{\bibfnamefont{J.}~\bibnamefont{Magueijo}} \bibnamefont{and}
  \bibinfo{author}{\bibfnamefont{L.}~\bibnamefont{Smolin}},
  \bibinfo{journal}{Class. Quant. Grav.} \textbf{\bibinfo{volume}{21}},
  \bibinfo{pages}{1725} (\bibinfo{year}{2004}), \eprint{gr-qc/0305055}.

\bibitem[{\citenamefont{Stelle}(1977)}]{Stelle:1976gc}
\bibinfo{author}{\bibfnamefont{K.~S.} \bibnamefont{Stelle}},
  \bibinfo{journal}{Phys. Rev. D} \textbf{\bibinfo{volume}{16}},
  \bibinfo{pages}{953} (\bibinfo{year}{1977}).

\bibitem[{\citenamefont{Horava}(2009)}]{Horava:2009uw}
\bibinfo{author}{\bibfnamefont{P.}~\bibnamefont{Horava}},
  \bibinfo{journal}{Phys. Rev. D} \textbf{\bibinfo{volume}{79}},
  \bibinfo{pages}{084008} (\bibinfo{year}{2009}), \eprint{0901.3775}.

\bibitem[{\citenamefont{Garattini and Saridakis}(2015)}]{Garattini:2014rwa}
\bibinfo{author}{\bibfnamefont{R.}~\bibnamefont{Garattini}} \bibnamefont{and}
  \bibinfo{author}{\bibfnamefont{E.~N.} \bibnamefont{Saridakis}},
  \bibinfo{journal}{Eur. Phys. J. C} \textbf{\bibinfo{volume}{75}},
  \bibinfo{pages}{343} (\bibinfo{year}{2015}), \eprint{1411.7257}.

\bibitem[{\citenamefont{Chatrabhuti et~al.}(2016)\citenamefont{Chatrabhuti,
  Yingcharoenrat, and Channuie}}]{Chatrabhuti:2015mws}
\bibinfo{author}{\bibfnamefont{A.}~\bibnamefont{Chatrabhuti}},
  \bibinfo{author}{\bibfnamefont{V.}~\bibnamefont{Yingcharoenrat}},
  \bibnamefont{and} \bibinfo{author}{\bibfnamefont{P.}~\bibnamefont{Channuie}},
  \bibinfo{journal}{Phys. Rev. D} \textbf{\bibinfo{volume}{93}},
  \bibinfo{pages}{043515} (\bibinfo{year}{2016}), \eprint{1510.09113}.

\bibitem[{\citenamefont{Channuie}(2019)}]{Channuie:2019kus}
\bibinfo{author}{\bibfnamefont{P.}~\bibnamefont{Channuie}},
  \bibinfo{journal}{Eur. Phys. J. C} \textbf{\bibinfo{volume}{79}},
  \bibinfo{pages}{508} (\bibinfo{year}{2019}), \eprint{1903.05996}.

\bibitem[{\citenamefont{Waeming and Channuie}(2020)}]{Waeming:2020rir}
\bibinfo{author}{\bibfnamefont{A.}~\bibnamefont{Waeming}} \bibnamefont{and}
  \bibinfo{author}{\bibfnamefont{P.}~\bibnamefont{Channuie}},
  \bibinfo{journal}{Eur. Phys. J. C} \textbf{\bibinfo{volume}{80}},
  \bibinfo{pages}{802} (\bibinfo{year}{2020}), \eprint{2005.08310}.

\bibitem[{\citenamefont{Faraoni}(2000)}]{Faraoni:2000wk}
\bibinfo{author}{\bibfnamefont{V.}~\bibnamefont{Faraoni}},
  \bibinfo{journal}{Phys. Rev. D} \textbf{\bibinfo{volume}{62}},
  \bibinfo{pages}{023504} (\bibinfo{year}{2000}), \eprint{gr-qc/0002091}.

\bibitem[{\citenamefont{Feng and Yang}(2017)}]{Feng:2017gms}
\bibinfo{author}{\bibfnamefont{Z.-W.} \bibnamefont{Feng}} \bibnamefont{and}
  \bibinfo{author}{\bibfnamefont{S.-Z.} \bibnamefont{Yang}},
  \bibinfo{journal}{Phys. Lett. B} \textbf{\bibinfo{volume}{772}},
  \bibinfo{pages}{737} (\bibinfo{year}{2017}), \eprint{1708.06627}.

\bibitem[{\citenamefont{Dehghani}(2020)}]{Dehghani:2020jcw}
\bibinfo{author}{\bibfnamefont{M.}~\bibnamefont{Dehghani}},
  \bibinfo{journal}{Eur. Phys. J. C} \textbf{\bibinfo{volume}{80}},
  \bibinfo{pages}{996} (\bibinfo{year}{2020}).

\bibitem[{\citenamefont{Dehghani}(2021)}]{Dehghani:2021civ}
\bibinfo{author}{\bibfnamefont{M.}~\bibnamefont{Dehghani}},
  \bibinfo{journal}{Int. J. Geom. Meth. Mod. Phys.}
  \textbf{\bibinfo{volume}{18}}, \bibinfo{pages}{2150046}
  (\bibinfo{year}{2021}).

\bibitem[{\citenamefont{Ling}(2007)}]{Ling:2006az}
\bibinfo{author}{\bibfnamefont{Y.}~\bibnamefont{Ling}}, \bibinfo{journal}{JCAP}
  \textbf{\bibinfo{volume}{08}}, \bibinfo{pages}{017} (\bibinfo{year}{2007}),
  \eprint{gr-qc/0609129}.

\bibitem[{\citenamefont{Mukhanov}(2005)}]{MukhanovBook}
\bibinfo{author}{\bibfnamefont{V.}~\bibnamefont{Mukhanov}},
  \emph{\bibinfo{title}{Physical foundations of cosmology}}
  (\bibinfo{publisher}{Cambridge University Press}, \bibinfo{year}{2005}).

\bibitem[{\citenamefont{Wu and Geng}(2012)}]{Wu:2011kh}
\bibinfo{author}{\bibfnamefont{Y.-P.} \bibnamefont{Wu}} \bibnamefont{and}
  \bibinfo{author}{\bibfnamefont{C.-Q.} \bibnamefont{Geng}},
  \bibinfo{journal}{Phys. Rev. D} \textbf{\bibinfo{volume}{86}},
  \bibinfo{pages}{104058} (\bibinfo{year}{2012}), \eprint{1110.3099}.

\bibitem[{\citenamefont{De~Felice and
  Tsujikawa}(2011{\natexlab{a}})}]{DeFelice:2011uc}
\bibinfo{author}{\bibfnamefont{A.}~\bibnamefont{De~Felice}} \bibnamefont{and}
  \bibinfo{author}{\bibfnamefont{S.}~\bibnamefont{Tsujikawa}},
  \bibinfo{journal}{Phys. Rev. D} \textbf{\bibinfo{volume}{84}},
  \bibinfo{pages}{083504} (\bibinfo{year}{2011}{\natexlab{a}}),
  \eprint{1107.3917}.

\bibitem[{\citenamefont{Sotiriou et~al.}(2011)\citenamefont{Sotiriou, Li, and
  Barrow}}]{Sotiriou:2010mv}
\bibinfo{author}{\bibfnamefont{T.~P.} \bibnamefont{Sotiriou}},
  \bibinfo{author}{\bibfnamefont{B.}~\bibnamefont{Li}}, \bibnamefont{and}
  \bibinfo{author}{\bibfnamefont{J.~D.} \bibnamefont{Barrow}},
  \bibinfo{journal}{Phys. Rev.} \textbf{\bibinfo{volume}{D83}},
  \bibinfo{pages}{104030} (\bibinfo{year}{2011}), \eprint{1012.4039}.

\bibitem[{\citenamefont{Li et~al.}(2011{\natexlab{b}})\citenamefont{Li,
  Sotiriou, and Barrow}}]{Li:2010cg}
\bibinfo{author}{\bibfnamefont{B.}~\bibnamefont{Li}},
  \bibinfo{author}{\bibfnamefont{T.~P.} \bibnamefont{Sotiriou}},
  \bibnamefont{and} \bibinfo{author}{\bibfnamefont{J.~D.}
  \bibnamefont{Barrow}}, \bibinfo{journal}{Phys.\ Rev.\ D}
  \textbf{\bibinfo{volume}{83}}, \bibinfo{pages}{064035}
  (\bibinfo{year}{2011}{\natexlab{b}}), \eprint{1010.1041}.

\bibitem[{\citenamefont{Wu}(2016)}]{Wu:2016dkt}
\bibinfo{author}{\bibfnamefont{Y.-P.} \bibnamefont{Wu}},
  \bibinfo{journal}{Phys. Lett. B} \textbf{\bibinfo{volume}{762}},
  \bibinfo{pages}{157} (\bibinfo{year}{2016}), \eprint{1609.04959}.

\bibitem[{\citenamefont{Golovnev and Koivisto}(2018)}]{Golovnev:2018wbh}
\bibinfo{author}{\bibfnamefont{A.}~\bibnamefont{Golovnev}} \bibnamefont{and}
  \bibinfo{author}{\bibfnamefont{T.}~\bibnamefont{Koivisto}},
  \bibinfo{journal}{JCAP} \textbf{\bibinfo{volume}{1811}}, \bibinfo{pages}{012}
  (\bibinfo{year}{2018}), \eprint{1808.05565}.

\bibitem[{\citenamefont{Maldacena}(2003)}]{Maldacena:2002vr}
\bibinfo{author}{\bibfnamefont{J.~M.} \bibnamefont{Maldacena}},
  \bibinfo{journal}{JHEP} \textbf{\bibinfo{volume}{05}}, \bibinfo{pages}{013}
  (\bibinfo{year}{2003}), \eprint{astro-ph/0210603}.

\bibitem[{\citenamefont{Amelino-Camelia
  et~al.}(2013{\natexlab{a}})\citenamefont{Amelino-Camelia, Arzano, Gubitosi,
  and Magueijo}}]{Amelino-Camelia:2013wha}
\bibinfo{author}{\bibfnamefont{G.}~\bibnamefont{Amelino-Camelia}},
  \bibinfo{author}{\bibfnamefont{M.}~\bibnamefont{Arzano}},
  \bibinfo{author}{\bibfnamefont{G.}~\bibnamefont{Gubitosi}}, \bibnamefont{and}
  \bibinfo{author}{\bibfnamefont{J.}~\bibnamefont{Magueijo}},
  \bibinfo{journal}{Phys. Rev. D} \textbf{\bibinfo{volume}{88}},
  \bibinfo{pages}{041303} (\bibinfo{year}{2013}{\natexlab{a}}),
  \eprint{1307.0745}.

\bibitem[{\citenamefont{De~Felice and
  Tsujikawa}(2011{\natexlab{b}})}]{DeFelice:2011zh}
\bibinfo{author}{\bibfnamefont{A.}~\bibnamefont{De~Felice}} \bibnamefont{and}
  \bibinfo{author}{\bibfnamefont{S.}~\bibnamefont{Tsujikawa}},
  \bibinfo{journal}{JCAP} \textbf{\bibinfo{volume}{1104}}, \bibinfo{pages}{029}
  (\bibinfo{year}{2011}{\natexlab{b}}), \eprint{1103.1172}.

\bibitem[{\citenamefont{Barrow and Cotsakis}(1988)}]{Barrow:1988xh}
\bibinfo{author}{\bibfnamefont{J.~D.} \bibnamefont{Barrow}} \bibnamefont{and}
  \bibinfo{author}{\bibfnamefont{S.}~\bibnamefont{Cotsakis}},
  \bibinfo{journal}{Phys. Lett. B} \textbf{\bibinfo{volume}{214}},
  \bibinfo{pages}{515} (\bibinfo{year}{1988}).

\bibitem[{\citenamefont{Barrow and Cotsakis}(1991)}]{Barrow:1991hg}
\bibinfo{author}{\bibfnamefont{J.~D.} \bibnamefont{Barrow}} \bibnamefont{and}
  \bibinfo{author}{\bibfnamefont{S.}~\bibnamefont{Cotsakis}},
  \bibinfo{journal}{Phys. Lett. B} \textbf{\bibinfo{volume}{258}},
  \bibinfo{pages}{299} (\bibinfo{year}{1991}).

\bibitem[{\citenamefont{Martin et~al.}(2014)\citenamefont{Martin, Ringeval, and
  Vennin}}]{Martin:2013tda}
\bibinfo{author}{\bibfnamefont{J.}~\bibnamefont{Martin}},
  \bibinfo{author}{\bibfnamefont{C.}~\bibnamefont{Ringeval}}, \bibnamefont{and}
  \bibinfo{author}{\bibfnamefont{V.}~\bibnamefont{Vennin}},
  \bibinfo{journal}{Phys. Dark Univ.} \textbf{\bibinfo{volume}{5-6}},
  \bibinfo{pages}{75} (\bibinfo{year}{2014}), \eprint{1303.3787}.

\bibitem[{\citenamefont{Sebastiani et~al.}(2014)\citenamefont{Sebastiani,
  Cognola, Myrzakulov, Odintsov, and Zerbini}}]{Sebastiani:2013eqa}
\bibinfo{author}{\bibfnamefont{L.}~\bibnamefont{Sebastiani}},
  \bibinfo{author}{\bibfnamefont{G.}~\bibnamefont{Cognola}},
  \bibinfo{author}{\bibfnamefont{R.}~\bibnamefont{Myrzakulov}},
  \bibinfo{author}{\bibfnamefont{S.~D.} \bibnamefont{Odintsov}},
  \bibnamefont{and} \bibinfo{author}{\bibfnamefont{S.}~\bibnamefont{Zerbini}},
  \bibinfo{journal}{Phys. Rev. D} \textbf{\bibinfo{volume}{89}},
  \bibinfo{pages}{023518} (\bibinfo{year}{2014}), \eprint{1311.0744}.

\bibitem[{\citenamefont{Kehagias et~al.}(2014)\citenamefont{Kehagias,
  Moradinezhad~Dizgah, and Riotto}}]{Kehagias:2013mya}
\bibinfo{author}{\bibfnamefont{A.}~\bibnamefont{Kehagias}},
  \bibinfo{author}{\bibfnamefont{A.}~\bibnamefont{Moradinezhad~Dizgah}},
  \bibnamefont{and} \bibinfo{author}{\bibfnamefont{A.}~\bibnamefont{Riotto}},
  \bibinfo{journal}{Phys. Rev. D} \textbf{\bibinfo{volume}{89}},
  \bibinfo{pages}{043527} (\bibinfo{year}{2014}), \eprint{1312.1155}.

\bibitem[{\citenamefont{Costa and Nastase}(2014)}]{Costa:2014lta}
\bibinfo{author}{\bibfnamefont{R.}~\bibnamefont{Costa}} \bibnamefont{and}
  \bibinfo{author}{\bibfnamefont{H.}~\bibnamefont{Nastase}},
  \bibinfo{journal}{JHEP} \textbf{\bibinfo{volume}{06}}, \bibinfo{pages}{145}
  (\bibinfo{year}{2014}), \eprint{1403.7157}.

\bibitem[{\citenamefont{Cai et~al.}(2014)\citenamefont{Cai, Gong, and
  Pi}}]{Cai:2014bda}
\bibinfo{author}{\bibfnamefont{Y.-F.} \bibnamefont{Cai}},
  \bibinfo{author}{\bibfnamefont{J.-O.} \bibnamefont{Gong}}, \bibnamefont{and}
  \bibinfo{author}{\bibfnamefont{S.}~\bibnamefont{Pi}}, \bibinfo{journal}{Phys.
  Lett. B} \textbf{\bibinfo{volume}{738}}, \bibinfo{pages}{20}
  (\bibinfo{year}{2014}), \eprint{1404.2560}.

\bibitem[{\citenamefont{Chakravarty and Mohanty}(2015)}]{Chakravarty:2014yda}
\bibinfo{author}{\bibfnamefont{G.~K.} \bibnamefont{Chakravarty}}
  \bibnamefont{and} \bibinfo{author}{\bibfnamefont{S.}~\bibnamefont{Mohanty}},
  \bibinfo{journal}{Phys. Lett. B} \textbf{\bibinfo{volume}{746}},
  \bibinfo{pages}{242} (\bibinfo{year}{2015}), \eprint{1405.1321}.

\bibitem[{\citenamefont{Kallosh and Linde}(2021)}]{Kallosh:2021mnu}
\bibinfo{author}{\bibfnamefont{R.}~\bibnamefont{Kallosh}} \bibnamefont{and}
  \bibinfo{author}{\bibfnamefont{A.}~\bibnamefont{Linde}}
  (\bibinfo{year}{2021}), \eprint{2110.10902}.

\bibitem[{\citenamefont{Dai et~al.}(2014)\citenamefont{Dai, Kamionkowski, and
  Wang}}]{Dai:2014jja}
\bibinfo{author}{\bibfnamefont{L.}~\bibnamefont{Dai}},
  \bibinfo{author}{\bibfnamefont{M.}~\bibnamefont{Kamionkowski}},
  \bibnamefont{and} \bibinfo{author}{\bibfnamefont{J.}~\bibnamefont{Wang}},
  \bibinfo{journal}{Phys. Rev. Lett.} \textbf{\bibinfo{volume}{113}},
  \bibinfo{pages}{041302} (\bibinfo{year}{2014}), \eprint{1404.6704}.

\bibitem[{\citenamefont{Munoz and Kamionkowski}(2015)}]{Munoz:2014eqa}
\bibinfo{author}{\bibfnamefont{J.~B.} \bibnamefont{Munoz}} \bibnamefont{and}
  \bibinfo{author}{\bibfnamefont{M.}~\bibnamefont{Kamionkowski}},
  \bibinfo{journal}{Phys. Rev. D} \textbf{\bibinfo{volume}{91}},
  \bibinfo{pages}{043521} (\bibinfo{year}{2015}), \eprint{1412.0656}.

\bibitem[{\citenamefont{Cook et~al.}(2015)\citenamefont{Cook, Dimastrogiovanni,
  Easson, and Krauss}}]{Cook:2015vqa}
\bibinfo{author}{\bibfnamefont{J.~L.} \bibnamefont{Cook}},
  \bibinfo{author}{\bibfnamefont{E.}~\bibnamefont{Dimastrogiovanni}},
  \bibinfo{author}{\bibfnamefont{D.~A.} \bibnamefont{Easson}},
  \bibnamefont{and} \bibinfo{author}{\bibfnamefont{L.~M.}
  \bibnamefont{Krauss}}, \bibinfo{journal}{JCAP} \textbf{\bibinfo{volume}{04}},
  \bibinfo{pages}{047} (\bibinfo{year}{2015}), \eprint{1502.04673}.

\bibitem[{\citenamefont{Panotopoulos et~al.}(2021)\citenamefont{Panotopoulos,
  Videla, and Lopez}}]{Panotopoulos:2020qzi}
\bibinfo{author}{\bibfnamefont{G.}~\bibnamefont{Panotopoulos}},
  \bibinfo{author}{\bibfnamefont{N.}~\bibnamefont{Videla}}, \bibnamefont{and}
  \bibinfo{author}{\bibfnamefont{M.}~\bibnamefont{Lopez}},
  \bibinfo{journal}{Eur. Phys. J. Plus} \textbf{\bibinfo{volume}{136}},
  \bibinfo{pages}{397} (\bibinfo{year}{2021}), \eprint{2001.05828}.

\bibitem[{\citenamefont{L\'opez et~al.}(2021)\citenamefont{L\'opez, Otalora,
  and Videla}}]{Lopez:2021agu}
\bibinfo{author}{\bibfnamefont{M.}~\bibnamefont{L\'opez}},
  \bibinfo{author}{\bibfnamefont{G.}~\bibnamefont{Otalora}}, \bibnamefont{and}
  \bibinfo{author}{\bibfnamefont{N.}~\bibnamefont{Videla}},
  \bibinfo{journal}{JCAP} \textbf{\bibinfo{volume}{10}}, \bibinfo{pages}{021}
  (\bibinfo{year}{2021}), \eprint{2107.07679}.

\bibitem[{\citenamefont{Amelino-Camelia
  et~al.}(2013{\natexlab{b}})\citenamefont{Amelino-Camelia, Arzano, Gubitosi,
  and Magueijo}}]{Amelino-Camelia:2013tla}
\bibinfo{author}{\bibfnamefont{G.}~\bibnamefont{Amelino-Camelia}},
  \bibinfo{author}{\bibfnamefont{M.}~\bibnamefont{Arzano}},
  \bibinfo{author}{\bibfnamefont{G.}~\bibnamefont{Gubitosi}}, \bibnamefont{and}
  \bibinfo{author}{\bibfnamefont{J.}~\bibnamefont{Magueijo}},
  \bibinfo{journal}{Phys. Rev. D} \textbf{\bibinfo{volume}{87}},
  \bibinfo{pages}{123532} (\bibinfo{year}{2013}{\natexlab{b}}),
  \eprint{1305.3153}.

\bibitem[{\citenamefont{Amelino-Camelia
  et~al.}(2013{\natexlab{c}})\citenamefont{Amelino-Camelia, Arzano, Gubitosi,
  and Magueijo}}]{Amelino-Camelia:2013gna}
\bibinfo{author}{\bibfnamefont{G.}~\bibnamefont{Amelino-Camelia}},
  \bibinfo{author}{\bibfnamefont{M.}~\bibnamefont{Arzano}},
  \bibinfo{author}{\bibfnamefont{G.}~\bibnamefont{Gubitosi}}, \bibnamefont{and}
  \bibinfo{author}{\bibfnamefont{J.~a.} \bibnamefont{Magueijo}},
  \bibinfo{journal}{Phys. Rev. D} \textbf{\bibinfo{volume}{88}},
  \bibinfo{pages}{103524} (\bibinfo{year}{2013}{\natexlab{c}}),
  \eprint{1309.3999}.

\end{thebibliography}

\end{document}